# Empirical and Experimental Perspectives on Big Data in Recommendation Systems: A Comprehensive Survey

[1*]Kamal Taha, [2]Paul D. Yoo, and [3]Aya Taha

**Abstract:** This survey paper provides a comprehensive analysis of big data algorithms in recommendation systems, addressing the lack of depth and precision in existing literature. It proposes a two-pronged approach: a thorough analysis of current algorithms and a novel, hierarchical taxonomy for precise categorization. The taxonomy is based on a tri-level hierarchy, starting with the methodology category and narrowing down to specific techniques. Such a framework allows for a structured and comprehensive classification of algorithms, assisting researchers in understanding the interrelationships among diverse algorithms and techniques. Covering a wide range of algorithms, this taxonomy first categorizes algorithms into four main analysis types: User and Item Similarity-Based Methods, Hybrid and Combined Approaches, Deep Learning and Algorithmic Methods, and Mathematical Modeling Methods, with further subdivisions into sub-categories and techniques. The paper incorporates both empirical and experimental evaluations to differentiate between the techniques. The empirical evaluation ranks the techniques based on four criteria. The experimental assessments rank the algorithms that belong to the same category, sub-category, technique, and sub-technique. Also, the paper illuminates the future prospects of big data techniques in recommendation systems, underscoring potential advancements and opportunities for further research in this field.

**Key words**: Big Data Algorithms, Recommendation Systems, Recommendation Algorithms, Deep Learning in Recommendations.

## 1 Introduction

The rapid advancement of big data has fundamentally transformed the analytical landscape, reshaping interactions and decision-making processes for both corporations and consumers [1, 2]. This transformation is fueled by the immense potential of big data, which enables users to explore a broad spectrum of questions, including unforeseen ones [3]. In today's digital age, personal mobile devices transcend their traditional role as communication tools; they are now potent data collectors, capturing every facet of an individual's digital footprint, from communications to online transactions. Corporations benefit from extensive data collection, gaining insights into individual and group preferences [4].

This torrent of data brings forth a significant challenge: the increasing complexity for users to make decisions that align with their personal needs [5]. Similarly, corporations grapple with the monumental task of processing and interpreting this data to anticipate user behaviors [6]. In response, recommendation systems have emerged as a pivotal solution within the realm of machine learning [7]. These systems use algorithms to analyze past user activities and their similarities with others, predicting future interests to provide personalized recommendations.

Recommendation systems have proliferated across various sectors, including Internet of Things (IoT) services, entertainment, e-learning, web search, bioinformatics, and engineering [8-11]. Their operational mechanisms range from content-based filtering, which builds user profiles based on item characteristics to recommend similar items, to collaborative filtering, relying on a user's and others' past behaviors, and to advanced deep learning-based systems that utilize AI algorithms to parse extensive datasets across platforms, adeptly personalizing user experiences [12]. Hybrid systems often yield more precise recommendations.

The role of data in shaping smart cities is also paramount [13, 14, 15]. This information is vital for smart city development, government planning, and enhancing individual comfort. Major companies like Amazon, Google, and Microsoft have effectively employed hybrid recommendation technologies to deliver personalized products and news.

Research in the field of recommendation systems can be traced back to the 1990s [16]. During this era, numerous heuristics were developed for content-based and Collaborative Filtering methodologies [17]. Gaining significant attention with the Netflix challenge, Matrix Factorization emerged as the predominant model in recommender systems from 2008 to 2016 [18, 19]. However, the linear characteristics of factorization models limited their effectiveness in handling large datasets, such as intricate user-item interactions and items with complex semantics [20, 21].

Coinciding with these developments, the mid-2010s witnessed a surge in the application of deep neural networks within machine learning, significantly impacting fields such as speech recognition, computer vision, and natural language processing [22]. This trend has also extended to graph models, where innovative multi-task prompting methods, inspired by prompt learning in NLP, have shown potential in enhancing the adaptability and effectiveness of graph-based machine learning applications, as demonstrated in recent studies [23].

The success of deep learning is attributed to the extensive expressiveness of neural networks, which are especially beneficial for learning from large datasets with complex patterns [24]. This advancement offered new prospects for enhancing recommendation technologies. Consequently, in recent years, there has been a notable increase in research focusing on neural network approaches to recommender systems [25]. Moreover, the integration of sociological behavioral criteria with data mining techniques, as noted in [26], highlights the complexity of user interactions in digital environments. Utilizing advanced models such as hypergraphs and neural networks, this approach is key to creating sophisticated and precise recommendation systems.

While several surveys have examined various facets of recommendation systems, this comprehensive survey aims to delve deeper into the current state of big data in recommendation systems, incorporating both experimental and empirical evaluations. Our goal is to provide an in-depth and nuanced understanding of this field, highlighting its achievements and challenges, thereby paving the way for future advancements.

[1] K. Taha is with the Computer Science Department, Khalifa University, Abu Dhabi, UAE. E-mail: kamal.taha@ku.ac.ae
[2] P. Yoo is with the Departmnent of Computer Science & Information Systems, University of London, Birkbeck College, UK, p.yoo@bbk.ac.uk
[3] A. Taha is with Brighton College, Dubai, United Arab Emirates, Email: bc004814@brightoncollegedubai.ae.

## 1.1 Motivation and Key Contributions

1) Main Challenge and Proposed Solution
   a) **Current Issue:** The current surveys examining algorithms in Recommendation Systems utilizing Big Data are deficient in two primary respects: first, there is an absence of a comprehensive and current overview; second, the classification of algorithms is excessively general and lacks precision (e.g., [17], [27], [83], [98], [99]). This shortfall leads to ambiguity in categorizing diverse algorithms and consequently results in the application of uniform metrics, which may yield inaccurate evaluations.
   b) **Proposed Solution:** We propose a twofold approach: first, we conduct an in-depth analysis of current algorithms for Big Data in recommendation systems; second, we introduce a methodologically sound taxonomy. This taxonomy offers a hierarchical and detailed classification of these algorithms, fostering a more accurate and systematic categorization.

2) Comprehensive Survey and Enhanced Assessment
   a) **Survey Goals:** We provide a survey of algorithms, focusing on those that use the *same* categories, sub-categories, techniques, and sub-categories.
   b) **Benefits of the Taxonomy:** Employing our taxonomy enhances the precision of assessments and comparisons of algorithms. It offers insights into their strengths and weaknesses, aiding future research.

3) Empirical and Experimental Evaluations
   a) **Empirical Evaluation:** The study includes a detailed empirical evaluation of diverse techniques used in Big Data for recommendation systems.
   b) **Experimental Evaluation:** This study ranks the algorithms that utilize same categories, sub-categories, techniques, and sub-categories experimentally.

## 1.2 Proposed Methodology-Based Taxonomy

We classify the algorithms for Recommendation Systems with Big Data into four broad categories based on their **'Analysis Methods'**. These categories are: (1) User and Item Similarity-Based Methods, (2) Hybrid and Combined Approaches, (3) Deep Learning and Algorithmic, and (4) Mathematical Modeling Methods. Each of these primary categories is divided into more specific sub-categories. These sub-categories are then broken down into various techniques that utilize the principles of their respective sub-category. Lastly, these techniques are further categorized into more detailed sub-techniques. Fig. 1 shows our methodology-based taxonomy. Our taxonomy offers the following benefits:

- Structured Presentation: This approach offers a systematic framework for displaying survey findings. The arrangement of related methodologies within a hierarchical layout aids readers in understanding the paper's logical progression.
- Extensive Inclusivity: This classification system encompasses a broad range of relevant techniques. Its tiered structure assists in pinpointing unexplored areas and potential research opportunities.
- Technique Analysis: By categorizing comparable methods together, this taxonomy facilitates the comparison of various research techniques. It highlights both their commonalities and distinctions, enabling an evaluation of their respective merits and limitations.
- Enhanced Replicability: This categorization improves the ability to replicate research by providing clear descriptions of the methodologies.

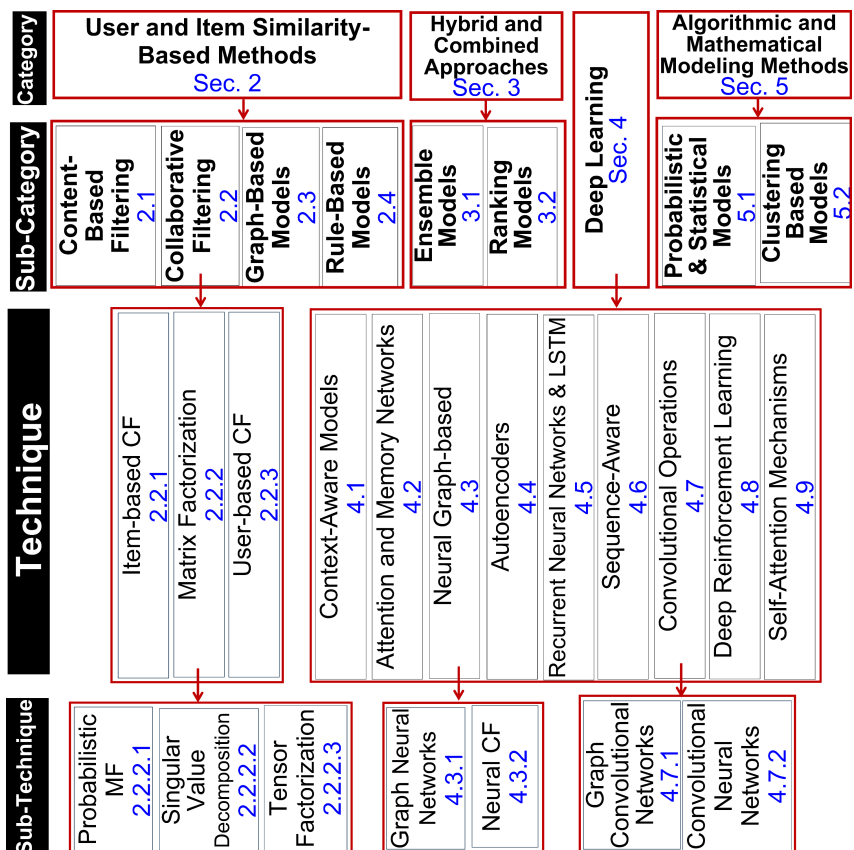

**Fig. 1:** This figure displays a structured, hierarchical approach to categorizing algorithms used in big data for recommendation systems. It methodically breaks down the algorithms into detailed categories, moving from broader methodology categories to more specific sub-techniques. The categorization in the top layer is based on analysis methods. The figure provides the corresponding section numbers in the manuscript.

## 2 User and Item Similarity-Based Methods

### 2.1 Content-Based Filtering

This approach to recommendations combines detailed analysis of item features with user preferences to provide personalized suggestions. Initially, it builds a user profile based on their interactions and preferences. This profile is then used to compare against the attributes of various items, which can be diverse and abundant in big data contexts, ranging from genre, keywords, and specific features relevant to the items, to more complex data like text descriptions, tags, categories, images, or audio features.

Items are recommended by scoring each one based on its similarity to the user's profile. For example, if a user shows a preference for movies by a certain director, the system will recommend other movies by that same director. The effectiveness of this approach is heightened when there is detailed, descriptive data about each item, allowing for a more accurate match to the user's preferences. This technique utilizes advanced data processing and machine learning methods, such as natural language processing and computer vision, to handle big data's complexity. This ensures recommendations are personalized, relevant, and accurate, thus enhancing user experience by aligning suggestions closely with individual interests.

*The rationale behind the usage of this technique can be summarized as follows*: (a) Personalization: It focusing on the preferences of individual users, which leads to a high level of personalization, (b) Item Independence: It does not rely on other users' behavior, which is particularly beneficial in situations where there is sparse data on user interactions, (c) Transparency: Recommendations can be easily explained by item features, (d) New Item Inclusion: It can recommend new items that have no previous user interactions, (e) Handling Big Data: It can efficiently manage vast and diverse item features, making it a scalable solution, and (f) Stability: The recommendations are stable over time as they are based on the user's consistent preferences. This is useful for users with niche or specific tastes.

*The conditions for the optimal performance of this technique can be summarized as follows*: (a) Effective processing of textual data through advanced NLP techniques is crucial. This involves understanding context, semantics, and nuances in language, (b) Employing robust machine learning algorithms that can learn from data, adapt to new information, and make accurate predictions is essential, (c) Tailoring recommendations to users, rather than making broad, generic suggestions. This involves understanding unique user preferences, and (d) Implementing a feedback loop where the system learns from the user's reactions to its recommendations. This helps in refining accuracy.

*The limitations of this technique can be summarized as follows*: (a) The approach leads to a lack of diversity in recommendations. It tends to recommend items similar to those a user has liked in the past, leading to a filter bubble where the user is not exposed to a broader range of options, (b) For new users, there's a lack of historical data to base recommendations on. This makes it difficult to predict preferences, (c) new items without any user interactions or ratings can be challenging to recommend, as the system has no data to assess their relevance to users, (d) its effectiveness is dependent on the quality of the metadata associated with the items, and (e) it may struggle with subjective user preferences due to its difficulty in capturing subtleties.

**TABLE 1:** RESEARCH PAPERS THAT HAVE EMPLOYED CONTENT-BASED FILTERING FOR BIG DATA IN RECOMMENDATION SYSTEMS

| Paper/year | Dataset | Interpretability | Scalability | Efficiency | Each of these papers uniquely combines techniques, such as using sentiment to weight credibility in reviews, or employing network analysis for academic citations or travel route suggestions, to enhance the relevance and accuracy of their recommendations. |
|---|---|---|---|---|---|
| | | | | | **Description** |
| Hu et al. [27] -2020 | Amazon Camera review | Fair | Acceptable | Acceptable | The authors created CISER, a system merging user profiling, reviewer credibility, and sentiment analysis. It has five parts: 1) Extracting key features using context and sentiment; 2) Scoring reviewers based on expertise and trust; 3) Mining user interests from review styles; 4) Assigning ratings to features based on sentiment; and 5) Recommending products using credibility-weighted sentiment. |
| Wang et al. [28]- 2022 | DBLP and APS | Acceptable | Unsatisfactory | Good | The authors introduced CNCRec, a citation recommendation framework that combines collaborative filtering and network representation learning. It's tailored for diverse academic networks, creating a paper rating matrix from an attributed citation network, incorporating paper attributes such as topics from titles and abstracts. CNCRec optimizes citation representation by uniting the citation network and paper attributes. |
| Jiang [29] - 2016 | Travelogues and photos | Good | Acceptable | Unsatisfactory | The authors developed a system for personalized travel sequence recommendations, leveraging travelogues and user-uploaded photos with tags and location information. It prioritizes entire travel sequences over single sites, using a 'topical package space' with details like tags, cost, time, and season. The system matches and suggests travel routes using user similar users' travel histories. |
| Lian et al. [30] -2018 | LBSN dataset | Fair | Fair | Good | The authors introduced ICCF, a framework for implicit feedback collaborative filtering, which improves parameter learning using coordinate descent. They combined ICCF with graph Laplacian matrix factorization to demonstrate the effectiveness of user features in assessing mobility similarity for location recommendations in a large Location-Based Social Network dataset. |
| Liu et al. [31]-2019 | Author Collected | Acceptable | Fair | Good | The authors developed a system to help undergraduates locate and work on research projects. It analyzes students' data, activities, and networks for relevance, connectivity, and quality. Utilizing data mining, social network analysis, and bibliometric analysis, the system recommends projects that align with students' qualifications. |

### 2.2 Collaborative Filtering

#### 2.2.1 Item-Based Collaborative Filtering

The technique of item-based recommendation focuses on recommending items to users by analyzing the similarity between items, utilizing metrics such as cosine similarity, Pearson correlation, or adjusted cosine similarity. This approach contrasts with user-similarity methods, as it compares items based on user-item interaction data like ratings or viewing history. This process helps in understanding how different items relate in terms of user preferences. For instance, if a user highly rates a movie, the system recommends other movies like it based on these similarity metrics.

Item-based recommendation excels in large environments due to its scalability and efficiency, particularly in handling sparse data, resulting in more consistent and accurate suggestions. With fewer items than users and more stable item relationships, it's easier to manage and more effective in big data contexts.

*The rationale behind the usage of this technique can be summarized as follows*: (a) Item-based collaborative filtering scales better in big data settings, as item-item matrices are smaller than user-based ones, aiding in managing resources, (b) It typically offers more precise recommendations by analyzing detailed item relationships, (c) This method efficiently addresses the sparsity in user-item interactions common in big data, (d) Its recommendations are more transparent, being easier to understand due to item similarities, and (e) It effectively accommodates new users by suggesting popular items, independent of their past interactions.

*The conditions for the optimal performance of this technique can be summarized as follows*: Efficient item similarity calculation using methods like cosine similarity, Pearson correlation, or Jaccard index is key, chosen based on data type. In big data, enhancing sparse item-user matrices with dimensionality reduction or data imputation improves recommendation quality. Scalability, critical for handling growing data, is often managed with distributed computing frameworks such as Apache Hadoop or Spark. Balancing personalization and diversity in recommendations ensures both similarity and variety. For new items/users with minimal interaction history, content-based filtering or hybrid models are effective.

*The limitations of this technique*: Scalability challenges in large datasets require more processing power and memory for similarity calculations, risking performance issues. Sparse user-item interactions in big data often lead to inaccurate recommendations. New items or users lack historical data, hindering accurate recommendation. Frequent recommendation of popular items overshadows niche ones, reducing diversity. Changes in user preferences/item features make recommendations from item-based collaborative filtering outdated.

**TABLE 2:** RESEARCH PAPERS THAT HAVE EMPLOYED ITEM-BASED COLLABORATIVE FILTERING FOR BIG DATA IN RECOMMENDATIONS

| Paper/year | Dataset | Interpretability | Scalability | Efficiency | Description |
|---|---|---|---|---|---|
| | | | | | These papers combine advanced techniques like Locality-Sensitive Hashing and transfer learning with in-depth user behavior analysis and data analysis tools to enhance recommendation accuracy, privacy, and personalization. |
| Yan [32]-2018 | WS-DREAM | Unsatisfactory | Good | Good | The authors improved the item-based Collaborative Filtering (ICF) method by adding the LSH technique, aiming for secure data publication and efficient integration in distributed service recommendations. Their proposed ICFLSH method assists recommender systems in offering services while maintaining user privacy. |
| Dai et al. [10] - 2020 | MovieLens and Netflix | Unsatisfactory | Unsatisfactory | Good | The authors developed CoFiToR, a three-stage transfer learning-based collaborative filtering framework that refines user preferences from general to specific, enhancing top-N recommendations. They also introduced a logistics recommendation method encompassing user tracking, choice evaluation, matrix-based rating analysis, habitual behavior reflection, and user information analysis via a model and decoder. |
| Sun [33] -2022 | GroupLens MovieLens | Unsatisfactory | Fair | Acceptable | The authors introduced a modified fuzzy adaptive resonance theory-based biclustering method for user similarity, focusing on shared items. The approach combines biclustering for local similarity and prediction with item-based collaborative filtering for global predictions, culminating in a merged, comprehensive final prediction. |
| Zan [34] -2023 | MovieLens | Unsatisfactory | Fair | Fair | The authors introduced a logistics service recommendation technique involving user interaction tracking, influenced choice evaluation, matrix-based rating analysis, habitual behavior reflection with a time interval matrix, and user information analysis through a model and decoder. |

### 2.2.2 Matrix Factorization

#### 2.2.2.1 Probabilistic Matrix Factorization

The technique employs a probabilistic approach to decompose large user-item interaction matrices, which typically represent user-item interactions such as user ratings for movies on a streaming platform. Each entry in these matrices corresponds to a user's rating or implicit feedback for a specific item. This probabilistic nature is crucial for handling the uncertainty and noise inherent in real-world data, as it assumes that these interactions are influenced by latent (hidden) factors related to both users and items, like genres, movie length, and user preferences. By learning latent features for both users and items and assigning probability distributions to these features instead of fixed values, the technique is able to make personalized and accurate recommendations even with sparse and noisy data. It tackles sparse data by predicting missing matrix entries, allowing for user rating predictions on uninteracted items. Its flexibility and scalability make it effective for handling vast and intricate datasets, ensuring efficient, personalized recommendations.

*The rationale behind the usage of this technique can be summarized as follows*: PMF excels in handling large, sparse user-item matrices common in big data, efficiently processing millions of users and items. It extracts meaningful latent features, scaling well for big data applications. PMF's probabilistic approach models uncertainty in user preferences and item attributes, useful in noisy, incomplete data scenarios. By reducing data dimensionality and capturing latent user-item relationships, PMF often outperforms simpler methods.

*The conditions for the optimal performance of this technique can be summarized as follows*: Optimizing PMF in recommendation systems is anchored in large, high-quality datasets for accurate user-item interaction learning. Selecting an appropriate number of latent features is essential to avoid underfitting or overfitting and manage computational complexity. Critical to this process is the tuning of hyperparameters such as learning rate and regularization. Also, enhancing PMF with other methods can significantly boost its effectiveness.

*The limitations of this technique can be summarized as follows*: PMF in large recommendation systems struggles with sparse data and faces scalability issues as datasets grow, demanding more computational resources. Its linear approach may not capture complex user-item relationships as effectively as non-linear models. PMF is challenged by new users or items without sufficient historical data. Its performance depends heavily on precise hyperparameter tuning, with a risk of overfitting if feature dimensions are too high or regularization is not well-adjusted. The interpretability of PMF's latent factors is limited.

#### 2.2.2.2 Singular Value Decomposition (SVD)

SVD starts by representing the data as a matrix, where rows could be users and columns could be items. This matrix is often sparse as not all users interact with all items. SVD helps to fill in these missing values by approximating the original matrix. The decomposition results in three matrices: a diagonal matrix of singular values and two orthogonal matrices. These singular values are key to understanding the strength of the latent features in the data. By truncating these values, SVD can reduce the dimensionality of the data, which helps in handling big datasets more efficiently. This reduced representation is then used to predict missing entries in the original matrix.

**TABLE 3:** RESEARCH PAPERS THAT EMPLOYED PROBABILISTIC MATRIX FACTORIZATION FOR BIG DATA IN RECOMMENDATIONS

| Paper/year | Dataset | Interpretability | Scalability | Efficiency | Description |
|---|---|---|---|---|---|
| | | | | | These papers employ advanced matrix factorization (e.g., matrix factorization and variational inference) and auxiliary information integration, focusing on privacy and optimization through methods like local disturbance of ratings and dual neural network-based hyperparameter generation. |
| Yi et al. [35] - 2019 | MovieLens, Douban-Book | Fair | Unsatisfactory | Good | The authors introduced a deep learning collaborative filtering approach, deep matrix factorization, incorporating auxiliary information. It applies two transformation functions to create latent factors for users and items from different inputs. They created an embedding for recommendation systems that predicts positive feedback and condenses sparse, high-dimensional data into compact, low-dimensional vectors. |
| Hu et al. [36]-2022 | Movielens, Jester | Unsatisfactory | Good | Good | The authors created RAP, a privacy-focused recommender system framework. It mixes private user ratings with public ones using local disturbance, allowing algorithms to use modified ratings without accessing private ones. Users employ de-perturbation for recommendations. |
| Yao et al. [37] -2021 | ProgrammableWeb | Unsatisfactory | Acceptable | Good | The authors created a mashup service recommendation system by adding implicit API correlation regularization to matrix factorization. They emphasize the importance of API characteristics and past interactions with mashups for future API predictions. They detailed the model's components and proposed methods to understand their approach's functionality. |
| Shen et al. [38]-2021 | MovieLens, GroupLens | Unsatisfactory | Fair | Good | The authors developed DVMF, a Bayesian recommendation framework using variational inference for enhanced optimization. The Parametric Inference Model, with dual neural networks, generates hyperparameters for latent factors, leading to the Variational MF Model. The method merges implicit feedback and user/item data into one-hot format using Implicit Feedback. |

*The rationale behind the usage of this technique can be summarized as follows*: (a) SVD identifies latent features in data, like genres in movies, improving accuracy in predicting user preferences, (b) SVD effectively handles sparse data in recommendation systems by estimating missing values, aiding in accurate predictions even with limited data, (c) Advances in algorithms and computing allow SVD to efficiently process large datasets, making it practical for big data applications, (d) SVD improves recommendation quality by revealing subtle user preferences not obvious in basic analyses, (e) Integral to collaborative filtering, SVD predicts user preferences for uninteracted items based on user-item interactions, and (f) Despite its complexity, SVD's output is interpretable, offering transparency and aiding in system refinement.

*The conditions for the optimal performance of this technique can be summarized as follows*: (a) Regularization is essential for controlling overfitting in sparse datasets, it helps SVD models generalize better, (b) Implementations must support parallel processing and distributed computing for large datasets, (c) Tuning parameters, like the number of latent factors, using methods like cross-validation, is crucial for SVD's effectiveness, (d) Developing strategies for new users or items with limited data is vital for system consistency, (e) SVD models should accommodate changing user preferences and item popularity over time, and (f) Ensuring the model's robustness to missing data through effective imputation techniques is important for accuracy.

*The limitations of this technique can be summarized as follows*: (a) SVD is ineffective with new users or items lacking historical data, (b) SVD may fail to capture complex, non-linear data relationships, (c) Performance heavily relies on correctly setting hyperparameters like latent factors, (d) SVD is prone to overfitting with datasets having numerous features, (e) Extracted latent features by SVD are often difficult to interpret, (f) The model's performance can degrade with noisy data, and (g) SVD does not naturally incorporate user-specific biases or preferences.

**TABLE 4:** RESEARCH PAPERS THAT HAVE EMPLOYED SINGULAR VALUE DECOMPOSITION FOR BIG DATA IN RECOMMENDATIONS

| Paper/year | Dataset | Interpretability | Scalability | Efficiency | Description |
|---|---|---|---|---|---|
| | | | | | These papers share techniques integrating advanced SVD and adaptive learning rates into recommender systems, aiming to improve recommendation accuracy, and optimizing algorithms for varied datasets and feedback using methods like semidefinite relaxation. |
| Guan et al. [39]- 2023 | Ciao, Epinions, Flixster | Fair | Acceptable | Acceptable | The authors introduced the Simultaneous Community detection and Singular Value Decomposition (SCSVD) framework, enhancing recommender systems via community detection. SCSVD combines community detection and recommendation model creation, linking ratings and social networks. It iteratively optimizes user preferences using social network communities, improving preference modeling. |
| Lian [40]- 2023 | Yelp and Amazon | Unsatisfactory | Fair | Good | The authors created a recommender system framework using discrete matrix factorization, accommodating diverse datasets and loss functions with explicit and implicit feedback, and auxiliary data. It uses block coordinate descent and semidefinite relaxation, functioning without hyperparameters in a two-phase item recall and ranking process. |
| Jiao [41]-2020 | Movielens | Unsatisfactory | Acceptable | Acceptable | The authors proposed an adaptive learning rate (ALR) function combining exponential and linear elements, integrated into the SVD++ recommendation algorithm. Featuring a high initial value, the learning rate quickly decreases in the middle phase and then slowly reduces to a smaller value at the end. |

#### 2.2.2.3 Tensor Factorization

The technique employs tensor factorization, a process that involves decomposing a high-dimensional tensor into multiple, lower-dimensional matrices or tensors. A tensor is essentially a multi-dimensional array of data. For example, in the context of a movie recommendation system, a 3D tensor might encompass dimensions for users, movies, and time. This setup represents how users' preferences for movies evolve over time.

Tensor factorization simplifies large, sparse tensors into smaller, dense ones, similar to dividing a complex puzzle into easier pieces. Techniques like Higher-Order Singular Value Decomposition or CANDECOMP/PARAFAC decomposition are used for this purpose. By reducing data dimensions, it makes data more manageable, focusing on retaining the most crucial features.

This approach is particularly crucial in big data contexts, where the volume of data can be overwhelming. Factorization helps to uncover latent (hidden) factors that influence user preferences and behaviors. For instance, in a movie recommendation system, these factors might represent underlying genres or themes that explain why certain users prefer certain movies. The goal is to discover these latent factors that explain observed data, thereby simplifying and enhancing the understanding of complex data patterns.

*The rationale behind the technique*: Tensor factorization enhances recommendation systems by handling multi-dimensional real-world data (like user, item, time), capturing complex patterns beyond traditional two-dimensional matrix factorization. This approach is especially beneficial in big data scenarios, where it can identify intricate relationships and offer more accurate predictions. It also allows for the inclusion of diverse side information and effectively manages sparse data. The multi-dimensional nature of tensors provides deeper insights into data and user behavior, improving recommendation accuracy.

*The technique's conditions for optimal performance*: Big data's high dimensionality and sparsity are addressed by PCA for dimension reduction and matrix completion for sparsity. Choosing correct hyperparameters like factors, regularization, and learning rates is essential for model success. Model complexity must be balanced to capture data patterns without overfitting, using cross-validation and regularization. Including contextual and temporal elements enhances performance. Testing various tensor factorization algorithms, such as CP and Tucker decomposition, finds the best fit for specific cases. Integration with other systems/data sources enhances recommendation efficiency.

*The limitations of this technique can be summarized as follows*: Tensor factorization struggles with scalability in large datasets due to increased computational complexity and memory needs. Sparsity in data, characterized by many missing tensor entries, hampers learning effective representations and predictions. It also faces the cold start problem, making it difficult to accurately recommend for new users or items with sparse data. To prevent overfitting and ensure better generalization in complex models, careful selection of rank and hyperparameters is crucial.

**TABLE 5:** FEATURING RESEARCH PAPERS THAT HAVE EMPLOYED TENSOR FACTORIZATION FOR BIG DATA IN RECOMMENDATIONS

| Paper/year | Dataset | Interpretability | Scalability | Efficiency | Description |
|---|---|---|---|---|---|
| | | | | | These papers use tensor factorization to uncover latent features and relationships in data, and the integration of auxiliary information, such as temporal, spatial, or graph-related data, to enhance prediction accuracy and address issues like data scarcity and cold-start problem |
| Ioannidis [42]-2021 | Digg datasets | Acceptable | Acceptable | Good | The authors proposed Coupled Graph-Tensor Factorization (CGTF), integrating graph-related auxiliary data into recommender systems. They designed an algorithm for factor matrix determination and missing entry completion. CGTF employs an ADMM solver for closed-form updates, supporting parallel and accelerated execution. This model also tackles the cold-start problem, occurring when tensors have incomplete slabs. |
| Li et al. [43]-2018 | QoS dataset | Unsatisfactory | Unsatisfactory | Good | The authors introduced a Time-aware Matrix Factorization model for cloud service QoS prediction, combining adaptive matrix factorization with temporal smoothing to provide accurate, time-sensitive forecasts for service recommendations. Utilizes temporal smoothing for more precise, timely QoS predictions in recommendations. |
| Meng [44]-2018 | QoS dataset | Acceptable | Good | Acceptable | The authors presented a service recommendation method combining time and location using tensor factorization. It uses CP decomposition for multi-dimensional analysis. It distinguishes between varying and stable QoS metrics, addresses data scarcity through time-region clustering, and uses CP decomposition for user-service-location relationships. |
| Yang [45]-2018 | Last.fm, Bibsonomy | Fair | Fair | Acceptable | The authors created Tagrec-CMTF, a tag recommendation system using CMTF, which leverages auxiliary matrices and tensor CP factorization to uncover latent features and optimize learning in tags, items, and users. It optimizes parameters for learning, reconstructs tensors, and exposes latent features for tags, items, and users. |

### 2.2.3 User-Based Collaborative Filtering

The technique in question involves creating detailed user profiles based on past behaviors and preferences, a key component of Big Data environments. This method operates on the assumption that users with similar tastes or behaviors in the past will likely continue to exhibit similar preferences in the future. To facilitate this, large datasets of user interactions and preferences, including ratings, views, purchases, or other forms of engagement with products, movies, or articles, are collected and analyzed. The technique uses metrics like cosine similarity to match user profiles with similar preferences, predicting a user's future likes based on others' tastes, enhancing recommendation accuracy.

*The rationale behind the usage of this technique can be summarized as follows*: It can effectively personalize recommendations by analyzing the behavior and preferences of similar users. It effectively uses implicit data, such as browsing history, to infer preferences, particularly when explicit feedback like ratings is limited. It diversifies recommendations by leveraging the varied tastes of a broad user base. It incorporates social relationships and community influences to tailor suggestions. It is effective in sparse big data environments.

*The conditions for the optimal performance of this technique*: (a) In big data, sparse user-item matrices are common. Using matrix factorization, dimensionality reduction, or adding more data sources, like content-based filtering, helps address this, (b) Recommendation systems need to balance personalization with diversity to prevent filter bubbles, thus broadening user exposure and fostering new interests, and (c) Incorporating both explicit (ratings) and implicit (click-through rates) user feedback is key for continually improving recommendation algorithms.

*The limitations are*: (a) It struggles with scalability in large systems, causing performance issues; (b) Sparse interactions in big datasets hinder finding similar users, reducing recommendation accuracy; (c) New users or items with limited history pose data insufficiency problems; (d) A bias towards recommending popular items often overlooks niche content; and (e) It tends to reinforce content homogenization, limiting user experience diversity.

**TABLE 6:** RESEARCH PAPERS THAT HAVE EMPLOYED USER-BASED COLLABORATIVE FILTERING FOR BIG DATA IN RECOMMENDATIONS

| Paper/year | Dataset | Interpretability | Scalability | Efficiency | Description |
|---|---|---|---|---|---|
| | | | | | These papers use social and behavioral theories (like Structural Balance Theory and emotional contagion) to understand user relationships and preferences, and the application of CF to process large datasets for personalized product or content recommendations. |
| Qi et al. [46]-2018 | MovieLens | Unsatisfactory | Unsatisfactory | Good | SBT-Rec is a method developed by the authors, utilizing Structural Balance Theory. It identifies a target user's "enemies" (users with contrasting preferences) in E-commerce, then finds "possible friends" (considering "enemy's enemy as a friend"). It recommends products liked by these "possible friends," which also match the user's preferences. |
| Wu [47]-2023 | Author Collected | Fair | Fair | Acceptable | The authors introduced a system for recommending products, utilizing a collaborative filtering algorithm and extensive data search capabilities. This system employs its filtering algorithm to sift through vast datasets during data processing, enhancing efficiency and optimizing the performance of the recommendation system. |
| Hong [48]-2018 | Author Collected | Unsatisfactory | Unsatisfactory | Fair | The authors proposed a method for recommending content to groups using social media data, considering social influence, emotional contagion, and conformity. They use an emotion-based model to identify emotional connections among users, which informs their group recommendations. Then, they are incorporated into group recommendations. |

## 2.3 Graph-Based Models

The approach involves leveraging graph structures to model and analyze complex relationships in large datasets, where data is represented as a graph consisting of nodes and edges. Nodes typically represent items, such as products, movies, or articles, and users, while edges signify relationships or interactions like purchases, ratings, or views. The primary concept here is the utilization of graph theory to capture and analyze these relationships, making it easier to explore connections, detect patterns, and make predictions based on the nature and strength of these connections.

Enhancing this approach, the use of graph algorithms such as shortest path, clustering, or graph neural networks enables the analysis of graphs to uncover patterns in interconnected data, crucial for accurate recommendations. These algorithms analyze paths and connections, allowing the system to provide personalized recommendations based on user preferences and item connections in the graph. This results in tailored, context-aware suggestions, demonstrating how the combination of graph theory and advanced algorithms is a potent tool for harnessing information in complex datasets.

*The rationale behind the usage of this technique can be summarized as follows*: (a) effectively capture complex relationships in big data, ideal for modeling users, items, and content, (b) be scalable, handling large data volumes and easily adapting to new data types in big data scenarios, (c) offer personalized suggestions by understanding user preferences and behaviors, (d) address the sparsity in big data's user-item interactions better than traditional matrix approaches, (e) enhance recommendations by incorporating context like time, location, and social connections, and (f) reveal hidden patterns in data, improving recommendation quality and supporting collaborative filtering

*The conditions for the optimal performance of this technique can be summarized as follows*: (a) Graph construction is crucial for performance; nodes and edges must represent relevant data relationships, like user-product interactions in e-commerce, (b) Recommendation systems need to be accurate, personalized, and diverse to meet varied user preferences, (c) Sparse interaction data can be addressed using embedding learning to capture latent node and edge features, (d) Graph algorithms must be efficient and tailored to data characteristics and recommendation goals, and (e) Hybrid methods combining content-based and collaborative filtering help tackle the challenge of new users or items with minimal interaction history.

*The limitations of this technique can be summarized as follows*: (a) as the complexity and size of data grow, models require more computational resources. Scalability becomes a critical concern, especially for algorithms not designed to handle large-scale data efficiently, (b) Many models rely on rich interaction data to make accurate predictions. In scenarios where interaction data is sparse, such as with new users or items, the model's effectiveness can be significantly hindered. This is a common challenge in recommendation systems, (c) New items or users with limited historical data pose a 'cold start' challenge. The model struggles to make accurate recommendations due to the lack of past interaction or behavioral data to learn from, and (c) Building and maintaining sophisticated models for varied and large datasets require substantial resources, including specialized expertise and computational power. This complexity often translates into higher costs and longer development times.

**TABLE 7:** FEATURING RESEARCH PAPERS THAT HAVE EMPLOYED GRAPH-BASED MODELS FOR BIG DATA IN RECOMMENDATIONS

These papers employ graph models and attention mechanisms to enhance accuracy by capturing dynamic interactions and utilize advanced techniques like federated learning and data fusion, integrating various data types for relevant and efficient recommendations.

| Paper/year | Dataset | Interpretability | Scalability | Efficiency | Description |
|---|---|---|---|---|---|
| Chen et al. [49] - 2021 | Amazon Musical Instruments | Fair | Fair | Acceptable | The authors presented TMER, utilizing a knowledge graph to model dynamic interactions between users and items over time. Simplifying the complexity of traditional neural networks, TMER captures user history and item relationships using attention mechanisms, enhancing both the accuracy and explainability of recommendations. The approach effectively integrates users' sequential behavior and broader contextual information, enhancing existing recommendation methods by emphasizing temporal and path-based analysis. |
| Qi et al. [50] - 2022 | PW dataset | Unsatisfactory | Unsatisfactory | Good | The authors introduced WAR (Web APIs Recommendation) to aid developers in discovering web APIs via keyword searches, streamlining app development. This method integrates planning, discovery, and API selection through data-driven techniques. Developers input keywords to find suitable web APIs. WAR uses a data graph with APIs as nodes and compatibility connections, creating subgraphs for each keyword set to outline app solutions. |
| Zhou et al. [51] - 2019 | MovieLens | Fair | Unsatisfactory | Fair | The authors developed a model merging cyber and social computing for large-scale group decision-making (LSGDM) in online service social recommendation systems. They used a graph model in scholarly big data to depict LSGDM, profiling academic network decision-makers. This model assesses researchers' academic performance, connecting them through collaborative network interactions. Their method includes a network partitioning algorithm to identify key experts, and an improved random walk algorithm with restart. |
| Peng [52] - 2022 | Comments Dataset | Unsatisfactory | Unsatisfactory | Acceptable | The authors presented a privacy-focused stock recommendation method using federated learning, training word embeddings with encrypted investment forum data. Their Federated Meta Embedding blends insights from private forums and public social media. They expanded this with Federated Graph Meta Embedding, adding graph-based modeling to integrate public and private. |
| Zhu et al. [53] - 2022 | User-commodity behavior | Unsatisfactory | Unsatisfactory | Fair | The authors introduced a data fusion method for recommending mobile commerce services in tree-based networks. It entails storing tree-node relationships with redundant data to fully separate tree-type nodes for accessibility. The main objective is to efficiently store and retrieve structured data, using a tree synchronization model to boost server data buffering and access on mobile devices for enhanced efficiency. |
| Song [54] - 2023 | Foursquare dataset | Fair | Unsatisfactory | Unsatisfactory | The authors created a precise Spatiotemporal Network-Based Recommender for suggesting Points of Interest (POIs). Their model prioritizes spatial and temporal data, incorporating essential elements for accurate POI recommendations. It features well-designed meta-paths that account for both temporal and spatial aspects. |
| Wang et al. [55] - 2023 | Douban Movie | Unsatisfactory | Unsatisfactory | Unsatisfactory | The authors presented a framework, combining multi-community clustering and equitable decision fusion. This framework has three main parts: community exploration, local recommendations, and equitable decision fusion. It starts by creating overlapping communities through random walks on a user-item bipartite graph to connect users with similar motivations. The next step includes item ranking algorithms for each community. |

## 2.4 Rule-Based Models

The technique combines rule-based systems and graph algorithms to enhance the effectiveness of recommendation systems. Rule-based systems operate on a set of predefined rules or criteria, which are developed based on user behavior, item attributes, and other relevant data points. These rules can be segmented for different user groups based on demographics, past behavior, or other relevant factors, using data mining and analysis techniques to uncover patterns and trends that inform rule creation.

The technique synergizes graph algorithms like shortest path, clustering, or graph neural networks with rule-based systems to create a robust recommendation system. These algorithms analyze graph interconnections, uncovering patterns that enable highly personalized, context-aware recommendations based on user preferences. This integration not only harnesses predefined rules based on user data and item attributes but also exploits the data graph's complex interconnections. Further enhanced by recommendation models like collaborative and content-based filtering, this approach ensures a comprehensive and effective strategy for recommendations.

*The rationale behind the usage of this technique can be summarized as follows*: (a) The models are simpler and more transparent than complex machine learning models, making them ideal for sectors like finance or healthcare where explaining recommendations is crucial, (b) They offer greater control and customization, as rules can be easily modified, making them suitable for dynamic environments, (c) They efficiently manage large, varied datasets in big data contexts, quickly filtering relevant data using predefined criteria, (d) They are stable and reliable, not relying on variable training data, and are apt for applications requiring consistent behavior over time, and (e) Rule-based models complement machine learning by handling well-understood, codifiable aspects, while machine learning deals with more complex data nuances.

*The conditions for the optimal performance of this technique can be summarized as follows*: (a) Base the system's rules on sound logic and a deep understanding of user behavior and preferences, (b) Use scalable and efficient algorithms and data structures to manage large data volumes without performance loss, (c) Ensure the system can process data in near-real-time, crucial for adapting to rapidly changing user preferences, (d) Personalize recommendations using individual user profiles and contextual information for relevance and effectiveness, (e) Offer a diverse set of recommendations to prevent overfitting to known preferences and introduce users to new items, and (f) Design the model to handle sparse data, ensuring accurate recommendations in big data environments.

*The limitations of this technique can be summarized as follows*: In big data contexts, rule-based systems face challenges with large data volumes, necessitating too many rules. More rules increase complexity and reduce agility, especially when adapting to new data patterns. These systems, limited by set rules, offer less personalization and are less adept at predicting preferences than machine learning models. They tend to overfit historical data and struggle to adapt to new trends or unforeseen user behaviors. Their inflexibility in novel situations and data types, along with a focus on short-term, rule-defined patterns, means they may overlook longer-term user interests. They risk creating a "filter bubble," showing users only content like their past interactions.

**TABLE 8:** FEATURING RESEARCH PAPERS THAT HAVE EMPLOYED RULE-BASED MODELS FOR BIG DATA IN RECOMMENDATIONS

| Paper/year | Dataset | Interpretability | Scalability | Efficiency | Description |
|---|---|---|---|---|---|
| | | | | | These papers utilize advanced data analysis methods (like collaborative filtering, fuzzy weighted-iterative methods, and enhanced Apriori algorithm) to identify patterns and make personalized recommendations. Each one incorporates user feedback and behavior analysis. |
| Varlamis et al. [56]-2023 | Electric power | Unsatisfactory | Good | Good | The authors designed a collaborative filtering approach where users with comparable profiles share best practices. This system is bolstered by a sensor network that gathers extensive data on behavior and environment. By analyzing this data, the system identifies user habits and the specific micro-moments (or cues) that initiate these habits. It offers rule-based, personalized recommendations, tailoring these based on the user's latest activities and the information stored in the knowledge base. |
| Sumathi [57] - 2020 | Multilevel datasets | Fair | Unsatisfactory | Acceptable | The authors introduced an enhanced fuzzy weighted-iterative method to address the shortcomings in applying association rules focused on user requests and the visualization of rule discovery. This approach begins by integrating user feedback with post-processing to leverage semantic information. Following this, they developed rule schemas designed to accommodate and predict complex rules based on user expectations. |
| Zhang [58]-2018 | T10I4D | Unsatisfactory | Unsatisfactory | Good | The authors created MCRS, a course recommendation tool that enhances the Apriori algorithm using distributed computing. MCRS efficiently detects enrollment trends by initially preprocessing data with Hadoop, then analyzing it with Spark to extract association rules. These insights are then transferred to MySQL via Sqoop, improving user feedback and search efficiency. |
| Liao et al. [59]- 2021 | Author Collected | Unsatisfactory | Fair | Fair | The authors designed a rule-based system that employs social network data and purchase behaviors to recommend fan pages. Their objectives encompassed user categorization, understanding the link between fan motivations and social network choices, analyzing fan page followers' online purchases, creating purchase-based profiles for enhancing e-business social network recommendations |

## 3 Hybrid and Combined Approaches

### 3.1 Ensemble Models

This integrated approach leverages the Ensemble technique, combining multiple recommendation algorithms or models to enhance the accuracy and reliability of recommendations. By merging various algorithms, it aims to utilize the strengths of each while addressing their individual weaknesses. A key aspect of this technique is diversity, employing different models like collaborative filtering, content-based filtering, and others in tandem. This ensures a more thorough analysis of data, capturing various dimensions of user preferences and behaviors.

The Ensemble approach excels in handling big data, characterized by complexity and high dimensionality. It outperforms single-model systems by integrating various algorithms, including machine learning techniques like decision trees, neural networks, and k-nearest neighbors. This method employs strategies such as voting and averaging for different tasks, and advanced techniques like boosting and bagging to refine model weighting, significantly improving the precision and robustness of its predictions and recommendations.

*The rationale behind this technique*: These models in recommendation systems enhance accuracy by merging diverse algorithmic strengths, addressing user preferences and item characteristics. They reduce overfitting by balancing individual model biases, ensuring more stable predictions. These models adeptly manage big data's complexity and heterogeneity, including varied data types. Their diversity also fortifies system robustness, compensating for individual model weaknesses, particularly in handling data anomalies. They effectively capture complex, non-linear relationships present in big data.

*The conditions for the optimal performance of this technique*: (a) Ensemble models improve by merging diverse models with different algorithms, architectures, and parameters, enhancing robustness in recommendations, (b) They need to balance complexity and simplicity to avoid overfitting or underfitting, ensuring effective pattern recognition, (c) In big data, these models must be scalable and efficient, using parallel processing and dimensionality reduction for large datasets, (d) Regularization is crucial to prevent overfitting and enhance model generalization on new data, and (e) Tuning hyperparameters for each model is vital to optimize performance on validation data.

*The limitations are*: Ensemble models, combining multiple predictive models, require substantial processing power and memory, especially with big data. They risk overfitting, performing well on training data but poorly on new data, if individual models are too complex or poorly tuned. Integrating these models into existing systems can be complicated and resource-intensive. Ensemble models struggle with generalizing across data types or in fast-changing big data environments.

**TABLE 9:** FEATURING RESEARCH PAPERS THAT HAVE EMPLOYED ENSEMBLE-BASED MODELS FOR BIG DATA IN RECOMMENDATIONS

| Paper/year | Dataset | Interpretability | Scalability | Efficiency | Description |
|---|---|---|---|---|---|
| | | | | | These papers involve the use of ensemble methods, combining multiple algorithms or approaches for enhanced prediction accuracy, and employing both supervised and unsupervised learning methods for analyzing and processing large-scale data. |
| Jain [60]- 2022 | Airline Dataset | Unsatisfactory | Fair | Fair | The authors analyzed customer reviews and ratings to predict product recommendations. They used a two-part approach: sentiment analysis of reviews using an LSTM model to assess sentiments about airline services, and evaluation of customer ratings for different aspects of these services. Combining these methods, they formed an ensemble to effectively predict airline recommendations. |
| Silva [61]-2021 | Iris, Glass, Sonar | Good | Unsatisfactory | Acceptable | The authors introduced methodologies for recommending classifier ensemble structures, using meta-learning for optimal settings of classifier type, number of base classifiers, and aggregation method. The aim is to efficiently establish a robust ensemble structure using well-known algorithms like kNN, DT, RF, NB, and LR as base classifiers. |
| Hammou [6]-2019 | Yahoo! Webscope | Unsatisfactory | Good | Fair | The authors developed a distributed group recommendation system on Apache Spark for large-scale data. It merges dimension reduction with supervised and unsupervised learning, tackling the curse of dimensionality, identifying user groups, and improving prediction accuracy. |
| Huang [62]-2019 | MovieLens | Unsatisfactory | Fair | Fair | The authors introduced AMRE, which includes agents, a reward-function, and a roulette system. Each agent is a recommendation algorithm with a reward value based on recommendation accuracy, affecting its retention. The reward function modifies this value, rewarding accuracy and penalizing errors. Agents with low reward values are replaced using the roulette system. |

## 3.2 Ranking Models

The integrated approach to delivering personalized, relevant, and timely recommendations in big data environments involves a sophisticated interplay of data analysis, machine learning, and user-centric methodologies. It hinges on the use of advanced machine learning and data analytics techniques to interpret and navigate the immense datasets characteristic of these environments. At the core of this process are sophisticated ranking algorithms that order items or services according to a user's preferences. These algorithms often utilize advanced methods like matrix factorization, deep learning, neural networks, and other pattern recognition techniques. The complexity of these algorithms can vary, ranging from simpler statistical models to more complex neural networks, each chosen based on the nature of the data and the specific requirements of the system.

In big data recommendation systems, machine learning models like decision trees, neural networks, and gradient boosting machines are essential. Continuous refinement through reinforcement learning, feedback loops, and A/B testing ensures their relevance and accuracy. Regular assessment with metrics like click-through and conversion rates helps adapt these systems to evolving user behaviors, maintaining their effectiveness in the dynamic big data landscape.

*The rationale behind the usage of this technique can be summarized as follows*: Traditional recommendation techniques struggle with the vast datasets typical in big data environments, but machine learning and AI-powered ranking models can process these efficiently, ensuring timely and relevant recommendations. These advanced models can detect complex, non-linear patterns in user behavior and item attributes, uncovering relationships that are too intricate for manual analysis or simpler algorithms. ranking models play a crucial role in personalizing user experiences by sorting items such as products, movies, and articles based on the user's past behavior and preferences. This enhances user satisfaction, as it allows users to find relevant content.

*The conditions for the optimal performance of this technique can be summarized as follows*: (a) Model Complexity: Balance is key; overly complex models risk overfitting and poor generalization, while too simple models may not capture data nuances, (b) Algorithm Selection and Tuning: Essential for optimal performance, as different algorithms have varying strengths and weaknesses depending on the data and task, (c) Handling Data Sparsity: Common in big data, effectively addressed using techniques like matrix factorization, embedding, or deep learning, (d) Models must efficiently manage the vast volumes of big data, often necessitating distributed computing and efficient algorithms, and (e) Mitigating bias in recommendations is vital for ethical and effective system performance.

*The limitations of this technique can be summarized as follows*: (a) Ranking models may unintentionally magnify existing biases in the training data, causing recommendations to be unfair or discriminatory, potentially favoring specific user groups or item types, (b) There's a danger of models overfitting to complex, high-dimensional training data, impairing their ability to generalize to new, unseen data and reducing real-world effectiveness, (c) Focusing on accuracy, ranking models might compromise on diversity and novelty, often suggesting items too similar to users' past choices, thereby limiting exposure to new or varied options, and (d) Incorporating ranking models into existing systems and workflows in complex big data ecosystems can be difficult, demanding technical effort.

**TABLE 10:** FEATURING RESEARCH PAPERS THAT HAVE EMPLOYED RANKING-BASED MODELS FOR BIG DATA IN RECOMMENDATIONS

| Paper/year | Dataset | Interpretability | Scalability | Efficiency | Description |
|---|---|---|---|---|---|
| | | | | | These papers integrate advanced ML models (such as deep neural networks and Bayesian personalized ranking algorithms) with domain-specific features (like passenger demand, road conditions, or semantic correlations from image features to enhance the effectiveness of recommendation systems. |
| Huang [63]-2019 | GPS trajectory | Fair | Unsatisfactory | Good | The authors developed a recommendation system using the wide and deep model, which combines wide linear frameworks and deep neural networks for improved passenger identification in taxis. This system mimics experienced taxi drivers by focusing on passenger demand, road conditions, and potential earnings. |
| Li et al. [64] - 2019 | Wisdom Tourist | Unsatisfactory | Fair | Acceptable | The authors introduced a hybrid recommendation system that merges the Hierarchical Sampling Statistics (HSS) model with the Multimodal Visual Bayesian Personalized Ranking (MM-VBPR) algorithm. This system uses the DCA model to extract semantic correlations from image features and integrates them into the VBPR model, enhancing recommendation effectiveness. |
| Wang [65]-2019 | Video dataset | Good | Good | Acceptable | The authors proposed the Category-aided Multi-channel Bayesian Personalized Ranking (CMBPR) method. It combines content-based recommendations with a multi-channel BPR technique, using category-aided sampling and multi-channel sampling for a more nuanced understanding of user preferences in short video categories. |
| Lak [66]-2021 | Gowalla | Fair | Unsatisfactor | Fair | The authors conducted a comparative study of various Bayesian personalized ranking algorithms, including NBPO and LightGCN, using multiple datasets. They focused on hyperparameter tuning to determine the most effective configurations for these recommendation algorithms. |
| He et al. [67]- | Amazon dataset | Acceptable | Fair | Unsatisfactory | The authors explored the integration of style features in personalized recommendations with their Style-aware BPR (SBPR) model. This approach uses a convolutional neural network to extract style features and a hierarchical gram matrix to capture stylistic elements, enhancing the representation of style in recommendation task. |

## 4 Deep Learning

### 4.1 Context-Aware Models

The technique represents a significant evolution from traditional, generic recommendation methods to a more personalized and dynamic approach. It excels in leveraging contextual information such as location, time, user preferences, and environmental conditions, which enhances the relevance and personalization of recommendations. This method employs big data analysis, utilizing large and complex datasets to identify patterns, trends, and user behaviors. This improves the accuracy and effectiveness of recommendations by processing and analyzing diverse data types, from structured data like purchase histories to unstructured social media interactions.

This technique enhances recommendation systems by integrating detailed user profiling with advanced machine learning algorithms. These algorithms continuously evolve, adapting to new data and refining recommendations based on user interactions and current context. This approach not only improves the system's accuracy and relevance over time but also creates a more intuitive and responsive user experience. Moving from generic recommendations to a nuanced, context-aware strategy, it significantly elevates user satisfaction and engagement, marking a transformative development in recommendation system design.

*The rationale behind this technique*: (a) The technique enhances personalization by considering current context factors like location, time, device, and user mood, alongside past behavior, (b) It offers more accurate, real-time recommendations compared to traditional systems that rely on historical data, adapting to changing user preferences, (c) Suited for big data's volume, velocity, and variety, it efficiently processes and adapts to diverse and rapidly generated data, and (d) It increases user engagement by offering recommendations that align with the user's immediate needs and preferences,

*The conditions for the optimal performance of this technique*: (a) The ability to understand and utilize context effectively is crucial. This means the system should adapt its recommendations based on changing contexts, such as time of day, user location, or current events, (b) The model should be capable of learning individual user preferences and adapt recommendations accordingly. This requires sophisticated machine learning algorithms that can evolve with user behavior, and (c) The algorithms should be transparent and avoid biases. There should be mechanisms to avoid reinforcing stereotypes.

*The limitations of the technique are*: (a) Context modeling is complex due to the diversity in user location, time, social settings, and preferences. Efficiently capturing and processing this varied data poses a significant challenge, (b) Big data often has sparse datasets with many unobserved context-user-item interactions, making accurate predictions difficult, (c) Context-aware systems struggle with new users or items that lack interaction data, making accurate recommendations challenging, and (d) There's a risk of bias in models, favoring certain items or users. Ensuring fairness in recommendations is essential yet difficult.

**TABLE 11:** FEATURING RESEARCH PAPERS THAT HAVE EMPLOYED CONTEXT-AWARE MODELS FOR BIG DATA IN RECOMMENDATIONS

| Paper/year | Dataset | Interpretability | Scalability | Efficiency | Description |
|---|---|---|---|---|---|
| | | | | | These papers integrate contextual data (such as local points of interest, weather, user activities, and social networks) and the use of advanced ML methods (like deep learning, latent factor models, and tensor factorization) to enhance the accuracy and personalization of recommendations. |
| Liu [68] - 2022 | BEIJING dataset | Unsatisfactory | Good | Good | The authors developed Hydra, a multi-modal transportation planning system using deep learning. This system is sensitive to local points of interest and weather, integrating routing systems with urban data in a dual-layer framework. Hydra customizes routes for various travel modes, adapting to urban settings and user preferences. |
| Motwani [69]-2023 | Patient Dataset | Fair | Fair | Good | The authors developed SPM for advanced patient monitoring. SPMR uses data from Ambient Assisted Living devices to monitor vital signs, symptoms, and activities, assessing and predicting patient health, and activating necessary services. It's compatible with technologies like IoT, ML, and AI, and integrates medical guidelines with patient data. |
| Gao [70]-2020 | Phoenix datasets | Unsatisfactory | Unsatisfactory | Fair | The authors developed N-PRA, a Latent Factor model-based algorithm, integrates social media contexts with user interests and dynamic POI popularity. It considers both POI types and geographic proximity, and factors in users' social networks, significantly improving the accuracy of personalized POI recommendations. |
| Chou [71]-2020 | CoMoDa, Frappe | Fair | Good | Acceptable | The authors developed presented DropTF, a fast tensor factorization method for context-aware recommendations using implicit feedback. DropTF excels by leveraging large volumes of unobserved data, showcasing the benefits of including all unobserved data in training a TF-based context-aware recommendation system. |

## 4.2 Attention and Memory Networks

The integration of sophisticated concepts such as Attention Mechanism, Memory Networks, and Contextual Awareness significantly enhances the accuracy and relevance of recommendations in big data environments. Attention networks play a crucial role by prioritizing key user behaviors or items. For instance, if a user frequently purchases certain types of books, the system will focus more on these preferences when suggesting new books. This selective focus on aspects of data most relevant to a user's preferences ensures that the recommendations are more aligned with individual interests.

Memory networks store and retrieve user data, enhancing personalization in big data contexts by leveraging past behaviors for tailored recommendations. Contextual Awareness further refines this by considering factors like time and location, ensuring recommendations align with current user circumstances. The combination of these networks with attention networks, which focus on key data points, enables efficient navigation of extensive data, resulting in precise, individualized recommendations that reflect both comprehensive analysis and unique user needs.

**TABLE 12:** RESEARCH PAPERS THAT HAVE EMPLOYED ATTENTION AND MEMORY NETWORKS FOR BIG DATA IN RECOMMENDATIONS

| Paper/year | Dataset | Interpretability | Scalability | Efficiency | Description |
|---|---|---|---|---|---|
| | | | | | These papers use attention mechanisms to accurately capture user preferences and behavior, and the integration of personalized, context-aware, and dynamic models for better user-item interaction understanding and prediction. |
| Zhang et al. [72]-2022 | Xing and Reddit | Fair | Unsatisfactory | Acceptable | The authors created A-PGNN, combining a Personalized Graph Neural Network (PGNN) and a Dot-Product Attention mechanism. PGNN captures individual user behavior graph structures, focusing on the user's role in node embedding updates, unlike standard GNN models. The Dot-Product Attention, inspired by Transformers, accurately represents the impact of past sessions on the current one. |
| Chiang et al. [74]-2023 | MovieLens, Pinterest | Unsatisfactory | Fair | Good | The authors developed a recommendation system that dynamically adapts to user preferences by analyzing user behavior and recent interactions. It features a contextual item attention module for tracking preference changes and item relevance, and a multi-head attention module for handling diverse preferences. The system's accuracy is enhanced by including time-related item information, allowing for a personalized item representation in context. |
| Liu et al. [75]-2020 | Patio/Lawn dataset | Fair | Fair | Good | The authors developed the AAMN model, which uses historical reviews and ratings with an attention mechanism to assess data relevance and create user and product profiles. The model excels in extracting crucial information from reviews and combines static and adaptive features through a non-linear fusion layer and a deep interaction layer for detailed insights into user-item interactions. |
| Yin and Feng [76]-2022 | Amazon & Taobao | Good | Unsatisfactory | Acceptable | The authors developed a framework with enhanced attention, notably an external attention method, to better understand attention mechanism correlations and simplify complex self-attention processes. Aimed at visual tasks, it reduces computational load and captures relationships between samples. The framework also includes a multi-head mechanism for better performance. |
| Wang [77]-2022 | MovieLens | Unsatisfactory | Fair | Fair | The authors developed a recommendation system for top N items, using attention-based Seq2Seq and Long Short-Term Memory to analyze user behavior and optimize input selection and output simulation. It focuses on balancing training loss for better recommendation list generation and performance. |

*The conditions for the optimal performance of this technique*: Attention and memory networks are better suited for Big data environments as they can effectively process and learn from large datasets. Attention mechanisms in these networks focus on parts of the data more relevant to user preferences, leading to personalized and accurate recommendations. Memory networks enhance this by using a memory component to store and retrieve information from past user interactions, making recommendations more contextually relevant. These networks are particularly adept at handling sequential data, like in movie or music recommendations, understanding patterns over time. They also effectively manage data sparsity by focusing on the most informative interactions. Scalable, they adapt and learn efficiently with the growing volume of data in big data environments.

*The conditions for the optimal performance of this technique are as follows*: (a) The attention mechanism must differentiate essential from non-essential data points to boost recommendation precision, (b) The memory network must manage substantial data and retrieve pertinent information when necessary, (c) Scalability is crucial for coping with the growing amount, speed, and diversity of big data, requiring effective algorithms and expandable infrastructure, (d) Personalized recommendations based on individual user data are vital, necessitating an understanding of user behavior, preferences, and context, and (e) The system must continually learn from new data and adjust its recommendations.

*The limitations are*: (a) Attention and memory networks need considerable computational power, limiting their suitability for real-time environments, (b) Their effectiveness suffers from data scarcity, with accurate recommendations dependent on both the amount and quality of user data, (c) These networks are prone to overfitting on non-diverse, affecting their performance on new data, (4) They face challenges with new users or items due to their reliance on extensive historical data for accurate predictions, and (5) These models risk learning and propagating biases from their training data, potentially leading to unethical recommendations.

### 4.3 Neural Graph-Based

#### 4.3.1 Graph Neural Networks (GNNs)

GNNs efficiently process data represented as graphs, comprising nodes (like users or products) and edges (indicating relationships). These networks excel in extracting features from graph structures by aggregating information from neighboring nodes, enabling them to understand complex node relationships and overall graph dynamics. Central to GNNs is the message passing mechanism, allowing nodes to update their information through interactions with adjacent nodes, revealing intricate data patterns. GNNs employ techniques like convolution and pooling to disseminate information across the graph. This aids in offering personalized, context-aware recommendations by considering user-item interactions and the user's network position, leading to precise predictions. GNNs also refine these recommendations by learning specific user and item characteristics.

*The rationale behind this technique*: GNNs excel in big data environments, adeptly modeling complex interactions between entities like users and products. They provide a deep understanding of data relationships. GNNs are particularly suited for graph-structured data common in big data scenarios such as social networks, enabling direct leveraging of this structure with less preprocessing. GNNs generate accurate results by considering data's relational context, identifying patterns that other methods might miss. They effectively tackle sparsity in big data, by propagating information through graph structures.

*The conditions for the optimal performance of this technique*: (a) Given the vastness of big data, GNN should be scalable and efficient. Techniques like graph sampling, parallel processing, and efficient memory management are essential to handle large-scale graphs, (b) Choosing the right GNN architecture (like Graph Convolutional Networks, Graph Attention Networks, etc.) and tuning parameters (like the number of layers, type of pooling, learning rate, etc.) is critical for optimal performance, (c) Regularization techniques (like dropout, weight decay) are important to prevent overfitting, (d) The training strategy, including the choice of loss function, optimization algorithm, and training/validation split, impacts GNN performance, and (e) Balancing personalization with diversity in recommendations is important to maintain user satisfaction.

*The limitations are*: (a) GNNs struggle with scalability in big data scenarios, as they require processing a large number of nodes and edges, increasing computational complexity, (b) GNNs with multiple layers can cause over-smoothing, where node features become too similar, leading to information loss and decreased performance in recommendation tasks, (c) The presence of heterogeneous data (varied types of nodes and edges) in big data poses a significant challenge in developing GNN architectures that can effectively utilize this diversity, and (d) GNNs may inherit and amplify biases from training data, making ensuring fairness and avoiding discrimination in recommendations a major challenge.

TABLE 13: RESEARCH PAPERS THAT HAVE EMPLOYED GRAPH NEURAL NETWORKS FOR BIG DATA IN RECOMMENDATIONS

| Paper/year | Dataset | Interpretability | Scalability | Efficiency | Description |
|---|---|---|---|---|---|
| | | | | | These papers use graph-based models to analyze user-item interactions and the incorporation of advanced methods like attention mechanisms and neural networks to enhance the understanding and representation of these interactions. |
| Cui et al. [78] - 2024 | MovieLens, LastFM | Unsatisfactory | Unsatisfactory | Acceptable | The authors proposed recommendation model combines Graph Diffusion with the Ebbinghaus Curve to understand user interactions and memory patterns. The model identifies crucial memory paths in user interactions through a unique graph diffusion method, facilitating a deeper analysis of these interactions. By integrating the Ebbinghaus Curve, they personalize the model to reflect individual user's evolving interests and memory behaviors. |
| Wang [79]-2023 | MovieLens, Last.FM | Fair | Good | Unsatisfactory | The authors introduced MI-KGNN, a knowledge graph-based recommendation model that links users and items. It advances node representation through multi-dimensional interaction analysis and a dual attention mechanism, allowing users and items to affect neighboring nodes' significance and highlight interaction patterns in the graph. |
| Sun et al. [25]-2023 | MovieLens, Douban | Good | Unsatisfactory | Good | The authors proposed Separated Graph Neural Recommendation model which is a graph-based system that divides interaction networks into user and item segments. It effectively processes varied data, merges different methods for coefficient calculation, and uses a three-level attention mechanism for enhanced feature fusion, leading to more effective and adaptable data propagation. |
| Sang [80]-2023 | YouTub, Yelp | Unsatisfactory | Unsatisfactory | Acceptable | The authors developed a model that uses a hierarchical heterogeneous graph neural network and adversarial training to improve user and item embeddings for recommendations. The model uses fake nodes to challenge its accuracy, adjusting to reduce reliance on extra regularization. |
| Xie [81]-2022 | DivMat | Unsatisfactory | Unsatisfactory | Good | The authors developed GraphDR, a neural network for recommendations, comprising a Diversified Preference Network linking various nodes, a Field-Level Heterogeneous Graph Attention Network for learning node representations, and Online Multi-Channel Matching for targeted item selection. |

### 4.3.2 Neural Collaborative Filtering (NCF)

At its core, NCF uses collaborative filtering, predicting user preferences based on past interactions and similarities among users or items. NCF integrates deep learning to capture the non-linear and complex relationships in user-item interactions, going beyond traditional matrix factorization methods. The technique employs a multi-layer neural network architecture that allows for learning user-item interaction functions from data, offering a more flexible approach compared to traditional collaborative filtering.

*The rationale behind this technique*: (a) NCF overcomes sparsity and high dimensionality in big data, unlike traditional methods like matrix factorization, by leveraging deep neural networks to manage sparse data and extract features from high-dimensional spaces, thus enhancing recommendation accuracy, (b) NCF, with its neural network architecture, introduces non-linearity to effectively capture complex patterns in user-item interactions, addressing the limitations of traditional collaborative filtering in understanding non-linear relationships, (c) NCF scales efficiently for big data using advanced optimization and neural network parallel processing, and (d) integrates extensive auxiliary data for enhanced user preference understanding.

*The conditions for the optimal performance of this technique*: (a) Choose neural network architecture that matches data complexity to prevent overfitting, (b) Optimize hyperparameters like learning rate and neuron counts for better performance, (c) Employ L2 regularization and dropout in complex models to generalize better, (d) Apply advanced techniques such as embedding layers for complex interactions, and (f) Consider ethics and mitigate bias in recommendation systems for fairness.

*The limitations are*: (a) NCF models struggle with scalability in large datasets, leading to high computational costs and processing times, (b) They face challenges with data sparsity and the cold start problem, making it hard to predict for new users/items, (c) Overfitting is common in NCF models, causing poor performance on unseen data, (d) These models lack transparency in decision-making, affecting user trust and satisfaction, € NCF models favor popular items, neglecting less popular ones and reducing recommendation diversity, (f) Their effectiveness is compromised in scenarios with limited historical data, and (g) Using deep learning in recommendation systems raises privacy concerns due to the need for detailed user data.

TABLE 14: RESEARCH PAPERS THAT HAVE EMPLOYED NEURAL COLLABORATIVE FILTERING FOR BIG DATA IN RECOMMENDATIONS

| Paper/year | Dataset | Interpretability | Scalability | Efficiency | Description |
|---|---|---|---|---|---|
| | | | | | The common techniques in these papers are the use of neural networks and contextual information in recommendation systems, with a focus on improving accuracy and relevance through collaborative filtering, content-based, and sequential models. |
| Rehman [82]-2023 | MovieLens | Acceptable | Unsatisfactory | Fair | The authors created the Weighted Context-based Neural Collaborative Filtering model, which improves Neural Collaborative Filtering by using weighted contextual information. This model evaluates user interactions with items in different contexts, assigning varying weights to these contexts based on their relevance. |
| Wu et al. [83]-2023 | Yelp | Fair | Unsatisfactory | Fair | The authors examined neural network-based recommendation models for accuracy, classifying them into collaborative filtering, content-based, and sequential types. They reviewed major studies and innovations in each group, provided insights, and discussed future research areas like basic concepts, model development, evaluation, and reproducibility in these systems. |

## 4.4 Autoencoders

Autoencoders, utilizing dimensionality reduction techniques, are highly effective in managing the complexity of big data. These techniques compress high-dimensional data into a more manageable form while preserving essential information. Autoencoders achieve this through a process that involves compressing data into a lower-dimensional space (encoding) and then reconstructing it back to its original high-dimensional space (decoding). This capability makes them particularly suitable for handling big data scenarios, like user-item interaction matrices in recommendation systems, which often contain sparse data.

Autoencoders, leveraging their neural network architecture, are adept in recommendation systems, capturing complex data relationships and generating detailed embeddings of users and items. These embeddings accurately reflect user preferences and item features, enhancing behavior and preference prediction. Through learning from user-item interactions, autoencoders improve collaborative filtering, identifying and utilizing patterns to recommend items matching user preferences. Their ability to record these patterns in a latent space is key for tailored recommendations, adapting to each user's unique behaviors and preferences. This capability makes autoencoders highly effective in personalized recommendation systems, notably enhancing the accuracy of predicting user preferences.

*The rationale behind this technique are*: (a) Autoencoders compress high-dimensional data into lower-dimensional, dense representations, ideal for recommendation systems with sparse data (many users and items, few interactions), facilitating the extraction of meaningful patterns, (b) They learn to encode data in a way that preserves relevant information, capturing underlying patterns in user preferences or item characteristics for accurate recommendations, (c) They excel in managing sparse data in recommendation systems, learning to reconstruct missing data or predict preferences with limited interaction data, (d) They adapt with variations like Variational Autoencoders and enhanced collaborative filtering, and (e) They capture non-linear relationships as neural networks, crucial for complex systems.

*The conditions for the optimal performance of the technique are*: (a) The architecture of the autoencoder, such as deep or variational autoencoders, significantly affects performance. Tuning the number of layers and neurons per layer to match data complexity is necessary, (b) The choice of optimization algorithm (like Adam or SGD) and learning rate is critical. A too high learning rate can miss the global minimum, while a too low rate may slow down convergence, (c) Optimizing batch size and the number of training epochs is important. Smaller batch sizes can improve generalization but may increase training time, (d) Autoencoders require substantial hyperparameter tuning due to their complexity and big data diversity for optimal performance.

*The limitations of this technique can be summarized as follows:* (a) Training autoencoders for recommendation systems is challenging due to sparse user-item interaction matrices, leading to difficulty in learning meaningful data representations, (b) Autoencoders face scalability problems with extremely large datasets, resulting in high training time and computational resource demands, (c) They have the risk of overfitting, especially with complex models and limited training data, (d) They struggle with providing recommendations for new users or items with minimal historical data, a significant issue in dynamic environments, (e) They are sensitive to data quality variations, affecting recommendation accuracy in big data environments.

**TABLE 15:** FEATURING RESEARCH PAPERS THAT HAVE EMPLOYED AUTOENCODERS FOR BIG DATA IN RECOMMENDATIONS

These papers use models like autoencoders (specifically Variational Autoencoders) and matrix factorization for analyzing and learning from user/item interactions and preferences. They address the cold start problem by integrating both individual behaviors and group preferences, and complex data representations

| Paper/year | Dataset | Interpretability | Scalability | Efficiency | Description |
|---|---|---|---|---|---|
| Selvi et al. [84]-2022 | Medhelp dataset | Unsatisfactory | Acceptable | Good | The authors presented a secure health data recommendation method using a stacked de-noising convolution auto-encoder–decoder and a modified blowfish algorithm for privacy. This approach organizes patient data with Hadoop and analyzes both explicit and implicit patient information in a two-way system. It combines these features for effective recommendations. |
| Dong et al. [85]-2020 | MovieLens, MovieTwe | Fair | Unsatisfactory | Fair | The authors developed the HCRDa method, integrating matrix factorization with autoencoder learning to analyze user/item ratings and learn low-dimensional representations, reducing time complexity. It includes user/item attributes to mitigate the cold start issue. The algorithm produces a prediction matrix by multiplying learned user and item features. |
| Nguyen &Cho [86]-2020 | Gowalla dataset | Fair | Unsatisfactory | Acceptable | The authors created the Variational Autoencoder Mixture Model (VAE-MM) to improve online behavior suggestions. It combines individual and group preferences using VAE. Individual preferences reflect past behaviors, while group preferences capture common interests. Enhancements like dropout layers and skip connections in the VAE increase its accuracy. |
| Drif et al. [87]-2020 | Amazon Review | Unsatisfactory | Unsatisfactory | Acceptable | The authors developed EnsVAE for recommender systems, blending multiple sub-recommenders with a VAE to convert user-item interactions into interest probabilities, increasing accuracy. It offers enhanced data representation and adapts to various user interactions. EnsGG, a form of EnsVAE, combines GRU-MF CF with GloVe-content based filtering. |
| Zeng et al. [88]-2021 | MovieLens | Unsatisfactory | Unsatisfactory | Fair | The authors created NCAR, an autoencoder framework for collaborative recommendations using implicit user trust. NCAR combines trust data and user-item interactions by embedding a user co-occurrence matrix and employing a neural recommendation process. It extracts this matrix from ratings, applies an autoencoder with correlation regularization for user embeddings. |

## 4.5 Recurrent Neural Networks (RNN) and LSTM

RNNs excel at processing sequential data due to their ability to remember past inputs, which is crucial for pattern analysis and prediction. Long Short-Term Memory networks (LSTMs), a specialized form of RNNs, are designed to learn long-term data dependencies. Their advanced architecture, featuring a cell state and three types of gates (input, output, and forget), allows them to manage information flow effectively.

In big data and recommendation systems, RNNs and LSTMs are highly effective. They handle large volumes of sequential user data, capturing temporal user behavior patterns for accurate, personalized recommendations. LSTMs are adept at understanding extended user behavior, crucial for distilling user preferences or item characteristics. Autoencoders complement RNNs and LSTMs by processing non-sequential data, like user profiles or item descriptions. They reduce high-dimensional data into simpler forms, facilitating easier handling by RNNs or LSTMs within a temporal framework.

*The rationale behind this technique are*: (a) RNNs and LSTMs excel in handling sequential data like browsing and purchase histories, essential for predicting user actions in recommendation systems, (b) LSTMs remember long-term user interactions, crucial for understanding current interests and predicting future preferences in recommendation, (c) RNNs and LSTMs continuously update their understanding of user behavior, adapting to changing preferences, beneficial for evolving big data scenarios, (d) The architecture of RNNs and LSTMs enables them to identify intricate patterns, leading to more personalized recommendations, (e) RNNs and LSTMs are adept at processing unstructured data like text and images, vital for recommendations.

*The technique's conditions for optimal performance*: (a) The complexity of the model should be balanced. Overly complex models might overfit, while overly simple models might underperform, (b) Carefully choose the sequence length to capture relevant temporal dependencies, (c) Optimize these parameters for stable and efficient training, (d) Use methods like grid search or random search to find the optimal set of hyperparameters for your model, (e) Employ techniques to handle sparsity in recommendation data, such as using embedding layers, (f) Implement a mechanism for model adaptation in response to user feedback and evolving data patterns (g) integrating RNNs/LSTMs with techniques like CNNs for improved feature extraction.

*The limitations of this technique are:* (a) RNNs and LSTMs are complex and computationally intensive, making scaling difficult for large-scale recommendation systems, (b) LSTMs struggle with very long sequences, affecting their ability to capture long-term user preferences, (c) RNNs and LSTMs perform poorly with sparse data and in cold start scenarios, (d) The sequential processing of RNNs and LSTMs may not efficiently adapt to the rapidly evolving data in recommendation systems, and (e) The opaque nature of RNNs and LSTMs makes it difficult to understand their decision-making.

**TABLE 16:** RESEARCH PAPERS THAT HAVE EMPLOYED REURRENT NEURAL NETWORKS FOR BIG DATA IN RECOMMENDATIONS

| Paper/year | Dataset | Interpretability | Scalability | Efficiency | Description |
|---|---|---|---|---|---|
| | | | | | These papers use RNNs for prediction, and the integration of graph-based approaches (such as graph attention networks and attributed random walks) to incorporate network structure and relationships. |
| Wang [22]-2022 | Beijing taxi dataset | Good | Fair | Unsatisfactory | The authors developed a method using a neural network with A* for improving Personalized Route Recommendation. This model combines an attention-based RNN for calculating costs from the starting point to potential destinations and a position-aware graph attention network for estimating costs to the destination. |
| Zhang [89]-2020 | DBLP dataset | Unsatisfactory | Acceptable | Unsatisfactory | The authors introduced a system for academic social network friend recommendations using scholars' academic attributes from digital libraries. Their method employs an attributed random walk approach and a graph RNN framework, focusing on both network structure and scholar attributes. |
| Zhao [90]-2016 | Microblogging data | Acceptable | Good | Acceptable | The authors proposed using social and e-commerce network data for product recommendations. They utilized recurrent neural networks for e-commerce feature representation and modified gradient boosting trees for social network feature embedding, applying matrix factorization for cold-start product recommendations. |
| Shen [91]-2020 | Boston dataset | Fair | Unsatisfactory | Unsatisfactory | The authors explored the connection between listing descriptions and pricing, creating a price recommendation system for competitive pricing of new listings. Their approach used feedforward networks, LSTM, and mean shift algorithms, to enhance price prediction accuracy. |

## 4.6 Sequence-Aware Models

The technique involves Sequential Pattern Mining, where frequent patterns in user behavior are identified. This involves identifying and extracting patterns from sequences of user activities or behaviors, which helps in understanding and predicting future preferences. Temporal Dynamics are crucial, as they consider the timing of user interactions and acknowledge that preferences evolve over time. Context-Awareness is another key element, focusing on the context of user actions like location/time to enhance the recommendations.

Sequence-aware models often employ predictive algorithms like Markov Chains, LSTM (Long Short-Term Memory) networks, or GRU (Gated Recurrent Unit) networks to forecast future user actions based on past sequences. By analyzing sequential data, these models offer highly personalized recommendations. For example, a music streaming service might suggest songs based on the sequence of previously listened tracks. Techniques like word2vec or BERT (Bidirectional Encoder Representations from Transformers) can be adapted to transform sequences of user interactions into vector space, making it easier to apply machine learning models.

*The rationale behind this technique are*: (a) The models predict user preferences based on action sequences, offering nuanced insights in complex scenarios, such as e-commerce, to discern immediate interests and purchase intentions, (b) These models excel at capturing changes in user preferences over time, useful for platforms like streaming services to understand evolving tastes, (c) The models enhance accuracy and user experience by analyzing user interaction sequences for highly personalized recommendations, (d) These models use interaction sequences in low user-item interaction settings to enhance recommendations, addressing data sparsity effectively, and (e) These models merge sequence data with context factors (time, location, device) for more relevant recommendations in data-rich settings.

*The technique's conditions for optimal performance*: (a) Employing advanced sequence modeling techniques such as RNNs, Long LSTM networks, or Transformer-based models can significantly improve the ability to capture temporal dynamics in user-item interactions, (b) Personalization and Context-Awareness: The model should be capable of offering personalized recommendations by understanding individual user sequences. It should also consider contextual information like time of the day, location, or specific user states, (c) Maintain a balance between leveraging known user preferences and exploring new items to enhance user experience, and (d) Ensure the model is resilient to anomalies and generalizes well to new data or users through extensive validation and testing.

*The limitations of this technique are:* (a) Models excel in short-term patterns but struggle with long-term user behavior due to diverse, prolonged interactions, (b) As data volume increases, these models become less efficient, demanding more computational resources and reducing scalability, (c) With limited historical data, especially for new users/items, models encounter the cold start problem, affecting recommendation accuracy, (d) Deep learning-based sequence-aware models often overfit on large, diverse datasets, leading to poor performance on unseen data, (e) The opaque nature of deep learning recommendations complicates scenarios requiring clear reasoning, and (f) Sequence-aware models find integrating and processing varied data types (text, images, videos) in big data challenging.

**TABLE 17:** FEATURING RESEARCH PAPERS THAT HAVE EMPLOYED SEQUENCE-AWARE MODELS FOR BIG DATA IN RECOMMENDATIONS

| Paper/year | Dataset | Interpretability | Scalability | Efficiency | Description |
|---|---|---|---|---|---|
| | | | | | The papers use advanced data processing, like counterfactual thinking and probabilistic analysis, combined with diverse models such as heuristic, learning-based, and reinforcement methods, to improve recommendation accuracy across various domains. |
| Chen [92]-2023 | MovieLens | Fair | Unsatisfactory | Unsatisfactory | The authors developed a data augmentation framework for sequential recommendation systems, employing counterfactual thinking in user behavior sequences. This framework includes seven models categorized into heuristic-based (four models), learning-based (two models), and reinforcement learning (RL) methods (one model). |
| Xu [93]-2019 | e-government | Fair | Acceptable | Good | They developed an e-government recommendation system combining probabilistic semantic analysis with collaborative filtering, focused on user and item attributes, and utilizing historical data for group formation and sequence mining. It uses sequence mining tailored to e-government needs. |
| Cheng [94]-2023 | TripAdvisor | Acceptable | Fair | Unsatisfactory | The authors introduced Seq2CASE, a weakly supervised framework for extracting aspect scores from review comments. This method addresses the challenge of the absence of ground truth data for these scores, simplifying complex comments to estimate aspect scores. Seq2CASE demonstrates a strong correlation with actual user feedback. |

## 4.7 Convolutional Operations

### 4.7.1 Graph Convolutional Networks

GCNs for Big Data Recommendation seamlessly merge graph-based data representation with the powerful capabilities of convolutional neural networks, tailored to meet the demands of large-scale and diverse data sets. This innovative technique transforms the way data is handled, making it particularly effective for personalized and context-aware recommendations.

GCNs revolutionize recommendation systems by representing users and items as interconnected nodes in a graph, capturing relationships through edges. This graph-based structure, combined with convolutional neural network operations, allows GCNs to process and learn from the network's topology effectively. GCNs excel in extracting rich feature representations from nodes, uncovering latent user preferences or item characteristics that raw data might not reveal. They aggregate information from neighboring nodes, enriching each node's context and enhancing recommendation quality. Also, GCNs incorporate both node and edge attributes, like user demographics or interaction types, making them highly effective and scalable for personalized recommendations in complex datasets.

*The rationale behind this technique are*: (a) GCNs excel in big data applications like social networks and e-commerce by modeling relationships in graph data, (b) They efficiently manage sparse data in recommendation systems by aggregating neighboring node information, aiding in predictions despite limited data, (c) GCNs enhance learning by incorporating additional data such as user demographics, crucial in big data for improved recommendations, (d) Capable of learning complex data representations, GCNs capture intricate user-item interactions for nuanced, personalized recommendations, and (e) By utilizing user-item graph structures, GCNs identify latent user behavior factors, boosting accuracy.

*The technique's conditions for optimal performance*: (a) The architecture of the GCN, including the number of layers, type of convolutional layers, and activation functions, needs to be carefully designed. Hyperparameter tuning, including learning rate, dropout rate, and regularization, is also essential for optimizing performance, (b) Employing graph sampling, mini-batch training, and efficient matrix operations are key strategies for managing large-scale data, (c) To prevent overfitting/ensure generalization, regularization techniques, dropout, and data augmentation methods should be utilized, and (d) Performance enhancement can be achieved by integrating GCNs with other models like matrix factorization, deep learning models (e.g., autoencoders), or other graph-based models.

*The limitations of this technique are:* (a) GCNs can struggle with scalability due to the computational complexity of convolutions over large graphs. Big data environments often involve very large user-item graphs, (b) GCNs might not perform optimally in sparse user-item interaction matrix as they rely on the graph structure for learning, and sparse data can lead to less informative graph structures, (c) Multiple graph convolutions can make node representations too similar, reducing accuracy, (d) GCNs have difficulty recommending for new users or items due to few connections, (e) GCNs, as deep learning models, lack clear explainability, and (f) GCNs are prone to performance issues in noisy big data environments.

**TABLE 18:** RESEARCH PAPERS EMPLOYED GRAPH CONVOLUTIONAL NETWORKS FOR BIG DATA IN RECOMMENDATIONS

| Paper/year | Dataset | Interpretability | Scalability | Efficiency | Description |
|---|---|---|---|---|---|
| | | | | | These papers use GCN for leveraging complex relational data, and the application of adversarial training, attention mechanisms, and specialized graph structures to enhance recommendation by understanding user-item interactions and attributes. |
| Yu et al. [5] - 2022 | LastFM, Douban | Acceptable | Unsatisfactory | Good | The authors introduced an adversarial framework with GCN. It features a GCN-based autoencoder for improving relation data by detecting intricate connections and reconstructing social profiles, clarifying each user's social relations. Also, they crafted a GCN-based social recommendation module sensitive to varying social relation strengths. This framework uses adversarial training in a Minimax game to amalgamate all components. |
| Yue et al. [12]-2023 | Movielens, Douban | Fair | Unsatisfactory | Good | The authors developed the Attribute-Fusing Graph Convolutional Network for recommendation systems, focusing on the role of attributes in representation learning. They employed an attention-based approach to combine user and item attributes into a unified representation within a <user, item, attribute> graph. A specialized Laplacian matrix was formulated for this graph. |
| Rang et al. [95]-2023 | Yelp and Beibei | Unsatisfactory | Unsatisfactory | Fair | The authors developed the MBHCR model, which utilizes a heterogeneous graph to analyze complex user behaviors through varied interactions. This model features a multi-behavior relational aggregator to identify common interaction patterns, addressing user sparsity issues. Additionally, it incorporates a behavior comparison learning enhancer to differentiate and highlight key aspects of primary and secondary user behaviors. |
| Liu [96]-2023 | Gowalla & Yelp | Fair | Unsatisfactory | Good | The authors presented a method for modeling multi-grained popularity features and high-order connectivity, focusing on diverse user preferences in popularity features. They created the JMP-GCF model, which utilizes popularity-aware embeddings and a joint learning strategy to analyze multi-level popularity signals. |
| Yang [97]-2023 | Amazon Musical | Good | Unsatisfactory | Fair | The authors developed the PGIR model that fuses enhanced graph convolution with review properties, aligning property data with review text for in-depth content analysis. This model focuses on complex user-item relationships and employs sophisticated graph convolution to capture dynamic user-item features and collaborative patterns. |

## 4.7.2 Convolutional Neural Networks

The integration of CNNs in recommendation systems has transformed the way data is analyzed, especially in sectors like e-commerce and media streaming. The architecture of CNNs is intricately layered, enabling them to automatically detect and learn relevant features from raw data. This feature of learning is crucial for making accurate recommendations.

CNNs have a multi-layered structure where each layer detects different data aspects, from basic patterns in initial layers to complex features in deeper ones. Key to this are convolutional layers that filter and process data, and pooling layers that focus on significant features, with non-linear functions like ReLU enhancing pattern recognition. CNNs excel in analyzing visual content, extracting characteristics such as color and texture, and processing user interaction data (clicks, views, purchase history) for in-depth understanding of user preferences. They're often combined with other neural networks, like RNNs, to merge diverse data types, boosting recommendation accuracy and making CNNs vital in personalizing user experiences.

*The rationale behind this technique are*: (a) CNNs excel in extracting complex features from high-dimensional data like images and videos, outperforming traditional algorithms in recommendations by identifying key patterns, (b) CNNs adeptly handle and interpret unstructured data such as images, text, and videos, making them invaluable in recommendation systems, (c) CNNs offer advanced understanding of context and spatial relationships in data, (d) CNNs' adaptability through transfer learning allows systems to use pre-trained models, reducing the need for new data, (e) CNNs boost collaborative filtering by leveraging auxiliary data to address cold start and data sparsity, thus enhancing recommendation accuracy, and (f) CNNs excel in recommending visual content, like fashion or movies, due to their proficiency in image and video analysis, ensuring relevancy.

*The technique's conditions for optimal performance*: (a) Selecting the appropriate number and types of layers (convolutional, pooling, fully connected), filter sizes, and network depth is critical. Integrating CNNs with other neural networks like RNNs enhance the understanding of sequential patterns in user behavior, especially for recommendation systems, (b) Optimizing parameters such as learning rate, batch size, and epochs is crucial for CNN performance. Advanced techniques like Bayesian Optimization can further refine this process, (c) Regularization methods like dropout, L1/L2 regularization, and data augmentation are vital in big data scenarios to avoid overfitting, and (d) Utilizing pre-trained models is advantageous, especially with limited labeled data. These models, already trained on large datasets, can be fine-tuned for specific recommendation tasks.

*The limitations of this technique are:* (a) CNNs are not ideal for processing non-visual, non-spatial data like user behavior or preferences in recommendation systems, (b) Due to their deep architectures, CNNs risk overfitting, especially with insufficient training data, which is often the case in specific recommendation tasks, (c) CNNs require substantial computational resources, making them impractical for organizations with limited hardware or in scenarios demanding quick updates, (d) CNNs face challenges in scaling for large datasets and lack flexibility due to their fixed architecture, (e) The "black box" nature of CNNs is a drawback in where understanding the rationale behind recommendations is crucial, and(f) CNNs struggle to make accurate recommendations for new users or items with minimal historical data.

**TABLE 19:** RESEARCH PAPERS THAT EMPLOYED CONVOLUTIONAL NEURAL NETWORKS FOR BIG DATA IN RECOMMENDATIONS

| Paper/year | Dataset | Interpretability | Scalability | Efficiency | These papers use hybrid models combining deep learning with other data sources (text, ratings, and auxiliary data) and the employment of specialized neural network structures (CNNs, attention mechanisms, and multilayer perceptrons) to enhance feature extraction and interaction analysis. |
|---|---|---|---|---|---|
| | | | | | **Description** |
| Liu et al. [21]-2021 | Amazon Product | Good | Unsatisfactory | Acceptable | The authors created the NCTR model, a hybrid neural network that combines text and rating data for item recommendation. NCTR uses a convolutional neural network to analyze text context and a fusion layer to integrate these features. Additionally, it employs multilayer perceptrons to manage the nonlinear interactions of combined item and user latent features, enhancing the accuracy. |
| Chen [98]-2022 | Douban and Yelp | Acceptable | Unsatisfactory | Fair | The authors created a recommendation system that combines a heterogeneous information network and deep learning. This system uses network embedding and deep learning to analyze auxiliary data and extract features from reviews. It employs an attention mechanism and dual parallel CNNs for processing user and item word vectors, utilizing multiple kernels in the first CNN layer. |
| Ahammad [99]-2023 | Yelp | Good | Acceptable | Unsatisfactory | The authors developed a hotel recommendation system combining user collaboration, preference analysis, and similarity-based recommendations. It utilizes collaborative filtering and classified data management, incorporating capsules with convolutional kernels for in-depth feature analysis. The system features two layers: the first converts images into 256 feature maps on a 6x6 grid, and the second layer is similar in function. |
| Chenxu [100]-2022 | Amazon & Yelp | Fair | Unsatisfactory | Fair | The authors introduced the AUR framework, combining an uncertainty estimator with traditional recommendation models. Using aleatoric uncertainty, it aims to improve niche item recommendations while preserving general performance. AUR was successfully integrated with Matrix Factorization, LightGCN, and VAE models. |

## 4.8 Deep Reinforcement Learning (DRL)

DRL algorithms in big data recommendation systems learn through trial-and-error by interacting with vast datasets of user preferences and behaviors. Central to these systems is an agent, usually an algorithm, that makes decisions based on the observed user data, aiming to maximize rewards. DRL adapts its recommendations continuously, enhancing personalization by understanding and predicting user preferences. It excels in sequential decision-making, focusing on long-term rewards and balancing the exploration of new recommendations with the exploitation of known preferences. This approach is vital in dynamic big data environments, allowing DRL to effectively adapt to constantly evolving user preferences and data, ensuring more relevant and engaging recommendations over time.

*The rationale behind this technique are*: (a) DRL thrives in changing environments, such as recommendation systems where user preferences and trends are always shifting. It continuously updates its learning based on new data, ensuring recommendations stay relevant, (b) DRL excels at identifying complex patterns, offering personalized recommendations for each user, even in large-scale scenarios, (c) DRL focuses on long-term user engagement, offering recommendations that enhance satisfaction and retention, (d) DRL effectively manages recommendation systems by strategically assessing the long-term impact of decision sequences, and (f) DRL algorithms balance exploring new options and leveraging known preferences, crucial for evolving and reinforcing user interests in recommendations.

*The technique's conditions for optimal performance*: (a) In DRL for recommendation systems, the reward function must focus on long-term user satisfaction and engagement, incorporating complex metrics beyond immediate clicks or views, (b) DRL algorithms need to be scalable for big data and stable for consistent learning. Techniques like experience replay, target networks, and distributed learning are crucial, (c) DRL models in recommendation systems should continuously learn and adapt to changing user preferences in dynamic environments, requiring online learning, (d) DRL must strike a balance between exploring new recommendations and exploiting known user preferences, using strategies like ε-greedy or Upper Confidence Bound, and (e) The complexity of DRL demands computational efficiency, achieved through algorithm optimization.

*The limitations of this technique are:* (a) DRL models are complex and need substantial computing resources, making them less suitable for real-time systems that require quick responses, (b) DRL faces challenges in recommendation systems with new items or users, as it depends heavily on extensive interaction data, (c) Developing effective reward functions for DRL is difficult, particularly in recommendation systems where rewards are indirect or delayed, (d) In DRL, finding a balance between introducing new recommendations and adhering to known user preferences is challenging. Too much exploration can irritate users, while too little can lead to suboptimal strategies, and (e) DRL models may perpetuate and intensify biases present in their training data, leading to fairness issues in recommendations.

**TABLE 20:** RESEARCH PAPERS THAT HAVE EMPLOYED DEEP REINFORCEMENT LEARNING FOR BIG DATA IN RECOMMENDATIONS

| Paper/year | Dataset | Interpretability | Scalability | Efficiency | Description |
|---|---|---|---|---|---|
| | | | | | These papers use deep reinforcement learning for adaptive, personalized recommendations and the integration of innovative algorithms (such as curriculum learning, similarity aggregation, and modified random walk algorithms) to optimize model training and improve the system's performance. |
| Huang et al. [13] - 2023 | Geolife, Chengdu | Fair | Fair | Unsatisfactory | The authors created FedDSR, a daily schedule recommendation system using deep reinforcement learning within a federated learning framework. They applied curriculum learning for better local optimization and generalization and introduced a similarity aggregation algorithm to refine model quality using locally uploaded parameters. The system operates on encrypted local devices, ensuring privacy. Incorporating curriculum learning simplifies training the deep reinforcement model. |
| Zhou [101] - 2021 | DBLP, ResearchGate | Fair | Unsatisfactory | Acceptable | The authors created a Hierarchical Hybrid Network using deep reinforcement learning and a modified random walk algorithm for analyzing large dataset correlations in recommendation systems. This system, tailored for collaborative scholarly big data projects, effectively evaluates complex multi-level entity interactions, showcasing the HHN's capability in intelligent routing and data analysis. |
| Wang et al. [102] - 2023 | New York Check-in | Acceptable | Fair | Acceptable | The authors created a deep interactive reinforcement learning framework for studying geo-human interactions, consisting of representation and imitation modules. The representation module converts geo-human interactions into embeddings ('state'), and the imitation module, acting as a reinforced agent, recommends the next POI based on these states, imitating user visit patterns. The model's success improves through feedback. |

## 4.9 Self-Attention Mechanisms

The self-attention mechanism in models is pivotal in understanding and processing user interaction data, such as clicks, views, and purchases. By focusing on different parts of the input sequence, the mechanism is adept at capturing the dependencies and relationships between various items in a dataset. This is particularly useful as it enables the model to not just consider individual items in isolation but to understand how one item in a user's history might influence their interest in another. Consequently, this approach allows the model to dynamically adapt to the input data, emphasizing the most relevant parts for making predictions or recommendations.

Self-attention enhances user experience in e-commerce, content streaming, and social media by accurately discerning user history for personalized recommendations. It contextually analyzes behavior, distinguishing between casual browsing and specific purchase intentions, and dynamically adjusts to align recommendations with current preferences and future behavior.

*The rationale behind this technique are*: (a) Self-Attention is adept at handling sequential data like user histories in recommendation systems. It processes sequences in parallel, unlike traditional models like RNNs or LSTMs, enhancing training and inference efficiency, (b) Self-Attention is effective in recognizing relationships between distantly placed items in a user's interaction history, essential for recommendation systems. It computes attention across the sequence, allowing full contextual consideration, (c) Inherent flexibility and scalability make Self-Attention ideal for big data environments with high volume, variety, and velocity. It adapts well to large datasets and complex user behavior patterns, and (d) It improves recommendation systems' accuracy by deeply understanding user preferences and behavior, leading to more precise, personalized recommendations.

*The technique's conditions for optimal performance*: (a) Self-attention mechanisms must balance user preferences over time with immediate interests, necessitating an adept attention design, (b) These models, complex in nature, risk overfitting, especially with large datasets. Employing regularization methods like dropout and early stopping is crucial, (c) The success of self-attention depends on hyperparameters like attention head count, embedding size, and learning rates. Precise tuning of these is key, (d) In big data contexts, adapting self-attention to user behavior changes and contextual shifts, possibly by adding temporal or contextual elements, is vital, and (e) While self-attention is effective, optimal performance often requires integration with other elements, such as CNNs for feature extraction or RNNs for sequence processing, customized to the recommendation system.

*The limitations of this technique are:* (a) Self-attention requires significant memory resources, as it involves storing and processing all pairwise interactions between elements in the input sequence. In big data contexts, where the input size can be extremely large, this can lead to excessive memory demands, (b) While self-attention is adept at capturing global dependencies, it may not always efficiently capture very long-term dependencies in sequences, which is important in recommendation systems where historical data plays a crucial role, (c) In recommendation systems, self-attention can underperform due to sparse data and cold start issues, (d) Self-attention risks overfitting in data-rich but pattern-repetitive environments, (e) Self-attention can amplify biases, reducing recommendation diversity and perpetuating existing patterns, and (f) Self-attention models are computationally intensive, increasing training time.

TABLE 21: RESEARCH PAPERS THAT HAVE EMPLOYED SELF-ATTENTION MECHANISMS FOR BIG DATA IN RECOMMENDATIONS

| Paper/year | Dataset | Interpretability | Scalability | Efficiency | Description |
|---|---|---|---|---|---|
| | | | | | These papers use self-attention mechanisms and attention-based models with diverse data for personalized recommendations. They capture user preferences and behaviors through neural networks, metric learning, and information fusion. |
| Meng et al. [103] - 2023 | MIND-small and Adressa-1week | Acceptable | Fair | Acceptable | The authors introduced GAINRec, a news recommendation model that leverages both personalized preferences and collective behavior patterns. It uses self-attention and cross-attention mechanisms to understand user preferences from news titles semantically, and a global transition graph to capture common behavior patterns. GAINRec combines these insights with a heterogeneous transition graph attention network, improving news recommendation accuracy and trustworthiness. |
| Sun et al. [104]-2019 | Yoochoose and Diginetica | Fair | Unsatisfactory | Good | The authors developed SANSR, a self-attention network for session-based recommendation. It uses attention mechanisms for parallel processing, identifying relevant items in a session and assigning weights for predicting the next item. It includes an embedding layer, self-attention blocks, and a feedforward network, concluding with incremental training in the prediction layer. |
| Wang et al. [105]-2019 | CAMRa2011 & Meetup | Good | Fair | Fair | The authors introduced SACML, combining neural networks with metric learning for recommendations. It uses self-attention to determine the significance of group member interactions, accumulating group preferences and applying collaborative metric learning for group recommendation tasks. |
| Yang et al. [106]-2022 | MIND news recommendation | Unsatisfactory | Unsatisfactory | Good | The authors proposed the MIAR method, using multichannel information fusion for user interest recommendations. It combines user-clicked news and title embeddings with candidate news. MIAR consists of an interactive framework (MIF) and a distributed framework with an interest activation module, refining user profiles and personalizing user representations. |

## 5 Algorithmic and Mathematical Modelling Methods

### 5.1 Probabilistic and Statistical Models

The approach to big data challenges employs machine learning algorithms, notably those based on Bayesian methods, to provide accurate and personalized recommendations to users. Central to this approach is the use of statistical inference, which infers user preferences and patterns from data, embracing the complexities and uncertainties inherent in big data. It frequently uses probabilistic methods, like Bayesian networks, probabilistic latent semantic analysis, and Markov decision processes, to manage the uncertainty in predicting user preferences. These methods estimate the likelihood of a user's preference for an item, based on their history and the item's characteristics.

In this approach, Statistical Learning methods like collaborative filtering and Singular Value Decomposition (SVD) analyze user behavior and preferences to predict interests. Additionally, advanced models incorporate contextual factors like time and location, refining recommendations. These algorithms adapt and improve over time, enhancing prediction accuracy in complex data environments.

*The rationale behind this technique are*: (a) Probabilistic models are adept at handling the uncertainty and complexity inherent in user behavior and preferences. Big data environments often involve noisy, incomplete, or inconsistent data, and probabilistic models can effectively manage these challenges by making probabilistic inferences about user preferences, (b) Statistical models allow for the personalization of recommendations. They can analyze large datasets to identify patterns and trends specific to an individual. This enables recommendation systems to offer more tailored suggestions, improving user satisfaction, (c) These models are adaptable to various data types like ratings and clicks, crucial in big data's diverse data landscape, (d) Leveraging historical data and user interactions, these models precisely predict preferences, leading to efficient, learning-based recommendations, (e) Proper regularization techniques enable these models to avoid overfitting in high-dimensional big data, maintaining generalizable and relevant recommendations, and (f) Probabilistic and statistical models integrate contextual details (time, location, etc.) into recommendations, enhancing their relevance and timeliness.

*The technique's conditions for optimal performance*: (a) The complexity of the model should be balanced. Complex models may overfit the training data and not generalize well to new data. Regularization techniques and cross-validation can help in managing this balance, (b) These models should account for changes over time in user preferences and item popularity. Incorporating temporal dynamics can improve recommendation quality, (c) Big data environments require models that can efficiently scale with data volume, using techniques like stochastic gradient descent, online learning, and distributed computing, (d) Probabilistic models, such as matrix factorization techniques, are effective in dealing with sparse datasets, where many user-item interactions are unknown, (e) Models need to personalize recommendations based on user profiles and contextual information, adapting to individual preferences and contexts, and (f) Addressing the challenge of making recommendations for new users or items with limited data through content-based filtering or hybrid models.

*The limitations of this technique are:* (a) These models are prone to overfitting, especially when the data is sparse but high in dimensionality (common in recommendation systems). Overfitting leads to models that perform well on training data but poorly on unseen data, (b) Recommendation systems often suffer from data sparsity issues. Probabilistic and statistical models might not perform well under these conditions as they rely on sufficient data to understand and predict user preferences accurately, (c) These models have difficulty dealing with new users or items (cold start problem) because they lack historical data to make accurate recommendations, (d) Input data biases can be amplified in the recommendations, posing issues in scenarios requiring fairness and diversity, (e) Selecting and tuning the right model is complex and resource-intensive, necessitating extensive experimentation, (6) Often designed for short-term preferences, these models may fail to capture and predict long-term user behavior, (7) Their effectiveness heavily relies on the quality and quantity of data, with poor or incomplete data reducing recommendation accuracy.

**TABLE 22:** FEATURING RESEARCH PAPERS THAT HAVE EMPLOYED PROBABILISTIC/STATISTICAL FOR BIG DATA IN RECOMMENDATIONS

| Paper/year | Dataset | Interpretability | Scalability | Efficiency | Description |
|---|---|---|---|---|---|
| | | | | | These papers highlight techniques that combine sentiment analysis, matrix factorization, clustering and DL to deeply understand user preferences and item features. They also emphasize real-time data analysis and sequential modeling to tailor recommendations to dynamic conditions and domain-specific needs. |
| Cui et al. [107] - 2023 | Amazon Beauty & LastFM | Acceptable | Acceptable | Good | The authors presented the Factors Mixed Hawkes Process (FMHP), an approach for incremental recommendation systems that analyzes key factors influencing event generation. FMHP transforms events into sequences driven by these factors, using the Hawkes process for modeling based on past events. It integrates various functions to evaluate different event intensities and employs an incremental updating algorithm, enabling real-time adjustments in event intensity, thus improving the system's adaptability to changing user interactions. |
| Liu et al. [24]-2023 | Amazon food dataset | Acceptable | Fair | Fair | The authors created SAMF, a recommendation system combining sentiment analysis, matrix factorization, topic modeling via Latent Dirichlet Allocation, and deep learning with BERT. They generate topic distributions from reviews to shape user and item matrices, forming a comprehensive user-item preference matrix for rating predictions. |
| Mei et al. [108] - 2020 | AIS dataset | Fair | Fair | Acceptable | The authors devised an innovative algorithm for recommending substitute ports for container ships. This method, distinct from standard adjacency matrix techniques, utilizes NLP to analyze AIS data. It creates sequential berth records and models ports as vectors in a spatial embedding. The algorithm proposes ports comparable to the originally planned but inaccessible ones and verifies these suggestions by examining sailing distances. |
| Zhu et al. [109]-2023 | Author collected | Unsatisfactory | Acceptable | Fair | The authors studied a specialized educational information system that uses algorithms to recommend personalized content. The system analyzes user history and tag attributes, incorporating data pre-cleaning for accuracy. It uses a clustering algorithm for more efficient and in-depth analysis. The study also applies collaborative filtering, enhanced with information entropy and standard deviation, to identify user similarities. |
| Xia et al. [110]-2023 | Taxi trajectory | Unsatisfactory | Acceptable | Good | The authors developed BiA*-ACO, a hybrid algorithm that efficiently identifies taxi routes in urban networks. It merges the Bidirectional A-star algorithm's cost estimation with the Ant Colony algorithm's heuristic, enhancing search efficiency. Additionally, it improves the Ant Colony algorithm's pheromone update rules by incorporating the best routes from each iteration. |

## 5.2 Clustering-Based Models

The approach to data segmentation and recommendation involves employing clustering algorithms, which are essential to the core of these models. These algorithms, including K-means, hierarchical clustering, and Density-Based Spatial Clustering of Applications with Noise, are selected based on the specific requirements and characteristics of the data. Their function is to divide the data into groups or clusters based on similarities, where each cluster represents a group of users or items sharing similar characteristics. For instance, one cluster might consist of users who prefer action movies, while another could include those who frequently purchase sports equipment. Feature selection is vital in this process, focusing on the most relevant data features for accurate, efficient clustering. The model then uses these clusters to generate tailored recommendations, considering both the cluster's common attributes and individual user preferences within that group.

*The rationale behind this technique are*: (1) Clustering algorithms, such as K-means, hierarchical clustering, and DBSCAN, can efficiently handle large datasets. They group similar items or users into clusters, reducing the complexity of the data. This makes it easier to process vast amounts, (2) By clustering similar items or users, these models can offer more personalized recommendations. For instance, a user can be recommended items that are popular within their cluster but might not be widely known outside of it. This approach is useful when niche items get recommended alongside popular ones, and (3) Clustering helps in uncovering hidden patterns and relationships in the data, leading to more accurate recommendations.

*The technique's conditions for optimal performance*: 1) The choice of clustering algorithm greatly impacts performance. Algorithms like K-Means, Hierarchical Clustering, or DBSCAN should be chosen based on the dataset's characteristics (size, density, dimensionality). The algorithm should be efficient and scalable to handle big data volumes, (2) Finding the optimal number of clusters is crucial, using methods like the Elbow Method, Silhouette Analysis, or Gap Statistics to avoid overfitting and underfitting, and (3) In big data, clustering models need to be dynamic and adaptable for rapid preference changes, enabling periodic cluster updates with low overhead.

*The limitations of this technique are:* (1) While clustering is effective for grouping similar items or users, it may not scale well with extremely large datasets in big data scenarios. As the number of users and items increases, computational complexity can become a significant challenge, (2) Clustering tends to group similar items or users together, which can lead to a lack of diversity in the recommendations. This homogeneity might not always align with the users' desire for varied recommendations, and (3) The effectiveness of clustering-based recommendations relies heavily on the quality of clusters. Poorly defined clusters result in inaccurate or irrelevant suggestions.

**TABLE 23:** RESEARCH PAPERS THAT HAVE EMPLOYED CLUSTERING-BASED MODELS FOR BIG DATA IN RECOMMENDATIONS

| Paper/year | Dataset | Interpretability | Scalability | Efficiency | Description |
|---|---|---|---|---|---|
| | | | | | These papers use graph clustering and deep learning for personalized content recommendations, incorporating time factors and community insights to adapt to evolving user interests, and employing GNNs and optimized clustering algorithms. They emphasize the importance of handling new users/items. |
| Rostami [112]-2022 | Author collected | Unsatisfactory | Fair | Good | The authors developed a two-phase hybrid food recommender system to overcome limitations like disregarding ingredients, time factors, and new users/items. The first phase involves graph clustering for content-based recommendations, and the second uses deep learning for user and food item clustering. The system also incorporates strategies to handle time-related and community factors. |
| Nie [1]-2023 | Movielens | Fair | Fair | Acceptable | The authors created the UCMF model for better cross-domain recommendations. It uses graph neural networks to fuse user data from multiple domains, improving user representation, especially in data-scarce situations. |
| Cui et al. [9]-2020 | MovieLens and Douban | Acceptable | Fair | Good | The authors proposed TCCF, a collaborative filtering algorithm, combines Time Correlation Coefficient (TCC) and optimized CSK-means clustering. It simplifies large data problems through clustering, grouping similar users for efficient recommendations. This includes an improved K-means algorithm using Cuckoo search optimization and a time factor to track evolving user interests. PTCCF further personalizes this approach, focusing on user preference patterns. |

# 6 Experimental Evaluations

## 6.1 Evaluation Methodology

The following methodology was utilized in conducting the experimental evaluations:

> - **Evaluating individual techniques:** After a comprehensive review of papers presenting algorithms employing a specific technique, we identified the paper with the greatest impact. The algorithm detailed in this influential paper was chosen as the representative for its respective technique. To determine the most significant paper among those reporting algorithms using the same technique, we considered various factors including its innovative contributions and recency.
> - **Ranking the techniques within the same sub-category**: Initially, the Mean Average Precision (MAP) for each algorithm was calculated. Next, the MAPs of the algorithms using the same technique were averaged to determine the MAP representative of that technique. Finally, within each primary sub-category, the techniques were ranked based on their computed average MAPs.
> - **Ranking the sub-categories within the same category:** First, the MAPs of the algorithms using the same sub-category were averaged to determine the MAP representative of that sub-category. Then, within each primary category, the sub-categories were ranked based on their computed average MAPs.
> - **Ranking the categories within the same method:** First, the MAPs of the algorithms using the same category were averaged to determine the MAP representative of that category. Then, within each primary method, the categories were ranked based on their computed average MAPs.

## 6.2 Evaluation Setup

In the process of tuning the hyper-parameters for the selected algorithms, our initial step involved experimenting with various parameter values for each algorithm. Subsequently, we selected the set of parameter values that enabled each algorithm to attain its optimal recommendation performance. We list below these values:

- Hu et al. [27]: min_df: 0.01 (Exclude terms that have a document frequency lower than 1%), max_df: 0.8 (Exclude terms that have a document frequency higher than 80%), ngram_range: (1, 2) (Include both unigrams and bigrams), max_features: 10000, lambda: 0.01, learning_rate: 0.001, epochs: 50, batch_size: 32.
- Yan et al. [32]: Neighborhood Size (k): k = 20, Rating Scale: 1 to 5, Minimum Number of Ratings: Minimum 5 ratings per item, Data Sparsity Threshold: At least 10 interactions per user, Implementation Specific Parameters: Batch size = 32, Epochs = 10, and Learning rate = 0.01.
- Yi et al. [35]: Number of Factors (K): K = 50, Regularization Parameters ($\lambda$) for Users ($\lambda\_u$): 0.1, Regularization Parameters ($\lambda$) for Items ($\lambda\_i$): 0.1, Learning Rate ($\alpha$): $\alpha$ = 0.02, and Maximum Iterations: 500.
- Guan et al. [39]: Tolerance: 0.0001, Random Seed: 42, Max Iterations: 1000, and Learning rate = 0.01.
- Ioannidis et al. [42]: Rank: 10, Regularization Parameters ($\lambda\_users = 0.1, \lambda\_items = 0.1, \lambda\_time = 0.1$), Learning Rate: 0.01, Max Iterations: 1000, Initialization Method: Random, Convergence Criterion $\epsilon$: 0.001, and Loss Function: RMSE.
- Qi et al. [46]: Number of Neighbors (k): k = 20, Threshold for Similarity: 0.5, and Minimum Number of Ratings: 5.
- Zhang et al. [72]: Embedding Size: 128, Attention Layers: 2, Attention Heads: 4, Key Size: 32, Value Size: 32, Learning Rate: 0.001, Batch Size: 128, Epochs: 20, Regularization Term: 0.0001, and Dropout Rate: 0.5.
- Wang et al. [79]: Number of Layers: 3, Hidden Units: 64, Embedding Size: 128, Learning Rate: 0.001, Epochs: 30, Batch Size: 100, Dropout Rate: 0.5, Regularization Factor: 0.0001, Activation Function: ReLU, Edge Dropout Rate: 0.1.
- Selvi et al. [84]: Number of Layers: 3 layers, Number of Neurons in Each Layer: (Input Layer - 100 neurons, Hidden Layer - 50 neurons, Output Layer - 100 neurons), Activation Function: ReLU, Learning Rate: 0.001, Batch Size: 128, Number of Epochs: 100, and Dropout Rate: 0.5.
- Wang et al. [22]: Number of Layers: 2 layers, Number of Units per Layer: 128 units per layer, Dropout Rate: 0.5, Learning Rate: 0.001, Batch Size: 32, Epochs: 20, Sequence Length: 50, and Embedding Dimension: 100.
- Yu et al. [5]: Number of Layers: 3, Activation Function: ReLU, Dropout Rate: 0.3, Learning Rate: 0.001, Weight Decay: 0.0005, Number of Epochs: 200, Early Stopping: 10 epochs without improvement, Neighborhood Sampling: 10.
- Sun et al. [104]: Number of Layers (L): 6, Model Dimension (D): 512, Number of Heads in Multi-Head Attention (H): 8, Inner Layer Dimension of Feed-Forward Networks (d_ff): 2048, Dropout Rate: 0.1, Positional Encoding Dimension: 512, Learning Rate: 0.0001, and Batch Size: 64.
- Rostami et al. [112]: Number of Clusters: 5, Initialization Method: k-means++, Maximum Iterations: 300, Convergence Tolerance: 0.0001, Random State: 42, N_init: 10.

We searched for publicly available codes for these algorithms. We could obtain codes for the following papers: [42][4], [50][5], [21][6], [72][7], [22][8], [27][9]. For the remaining representative papers, we developed our own implementations using TensorFlow, as described by Sinaga and Yang [113]. We trained these implementations using the Adam optimizer, as suggested by Sinaga and Yang [113]. TensorFlow's APIs provide users with the flexibility to create their own algorithms [114]. Our development language was Python 3.6, and we utilized TensorFlow 2.10.0 as the backend for the models.

## 6.3 Compiling Datasets for the Evaluations

MovieLens dataset[10]: It is a repository of movie ratings. It integrates a vast array of user ratings along with movie metadata like titles, genres, release years, and user demographics. The 20M version, which we focus on, is substantial, featuring 20 million ratings and 465,000 tag applications for 27,000 movies by 138,000 users.

Amazon Electronics dataset[11]: It offers user reviews and detailed product information from Amazon. It encompasses an extensive array of electronic product reviews, featuring user ratings, textual feedback, and helpfulness votes. It includes product metadata descriptions, categories, pricing, brands, and image features. It encapsulates around 35 million reviews up to March 2013.

## 6.4 The Experimental Results

We ran the algorithms using an Intel(R) Core(TM) i7-6820HQ processor at 2.70 GHz, with 16 GB RAM, operating on Windows 10 Pro. We utilized the Mean Average Precision (MAP) metric for the evaluations. MAP calculates the average precision across different ranks, emphasizing the order of retrieved items. Precision at K (P@K) gauges the ratio of relevant items in the top K results.

---

[4] https://github.com/bioannidis/Coupled_tensors_graphs
[5] https://github.com/qlyseven/source-code
[6] GitHub - luojia527/NCTR_master: python
[7] https://github.com/CRIPAC-DIG/A-PGNN
[8] https://github.com/bigscity/NASR
[9] https://github.com/facebookresearch/fastText
[10] https://grouplens.org/datasets/movielens/
[11] https://www.kaggle.com/datasets/saurabhbagchi/amazon-electronics-data

The results are shown in Table 24 and Figs 2 and 3 as follows:
- Table 24 shows the overall average MAP scores of the selected algorithms. The table also shows the following: (1) the ranking the techniques within a same sub-category, (2) the ranking the techniques within a same sub-category, (3) the ranking the categories within a same method, and (4) the ranking of the methods.
- Figs. 2 and 3 illustrate the individual MAP@10, MAP@30, and MAP@50 for each algorithm using MovieLense dataset (Fig. 2) and Amazon Electronics Dataset (Fig. 3). The algorithms in each figure are grouped based on the common methods they employ.

**Table 24:** The overall average MAP of each selected algorithm Using MovieLense dataset (ML) and Amazon Electronics Dataset (AE). The table also shows the following: (1) the ranking the techniques within a same sub-category, (2) the ranking the sub-categories within a same category, (3) the ranking the categories within a same method, and (4) the ranking of the methods.

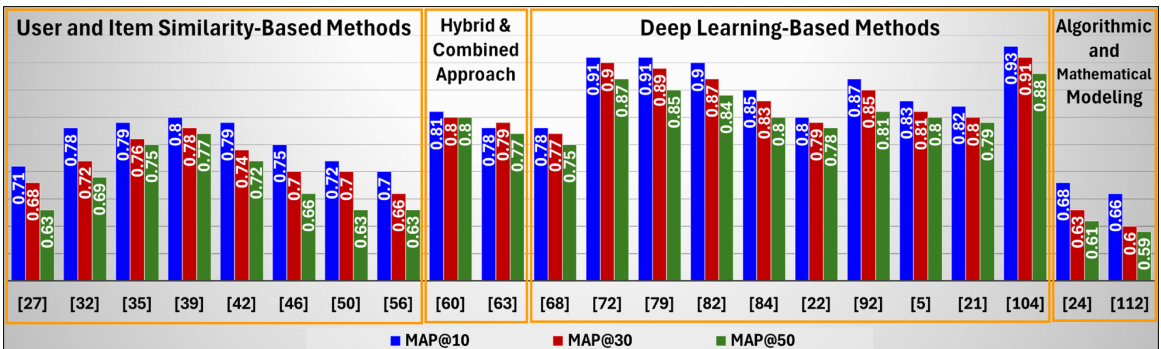

| Method | Category | Sub-Category | Technique | Paper | Data Set | Avg MAP | Tech Rank | Sub-Cat. Rank | Categ Rank | Method Rank |
|---|---|---|---|---|---|---|---|---|---|---|
| User and Item Similarity-Based Methods | Content-Based Filtering | N/A | N/A | [27] | ML / AE | 0.67 / 0.65 | N/A | N/A | 3 | 3 |
| | Collaborative Filtering | Item-based CF | N/A | [32] | ML / AE | 0.73 / 0.69 | N/A | 2 | | |
| | | Matrix Factorization | Probabilistic MF | [35] | ML / AE | 0.77 / 0.72 | 2 | 1 | 1 | |
| | | | Singular Value Decompositio | [39] | ML / AE | 0.78 / 0.74 | 1 | | | |
| | | | Tensor Factorization | [42] | ML / AE | 0.75 / 0.72 | 3 | | | |
| | | User-based CF | N/A | [46] | ML / AE | 0.70 / 0.68 | N/A | 3 | | |
| | Graph-Based Models | N/A | N/A | [50] | ML / AE | 0.68 / 0.66 | N/A | N/A | 2 | |
| | Rule-Based Models | N/A | N/A | [56] | ML / AE | 0.66 / 0.63 | N/A | N/A | 4 | |
| Hybrid & Combined Approach | Ensemble Models | N/A | N/A | [60] | ML / AE | 0.80 / 0.75 | N/A | N/A | 1 | 2 |
| | Ranking Models | N/A | N/A | [63] | ML / AE | 0.78 / 0.73 | N/A | N/A | 2 | |
| Deep Learning | Context-Aware Models | N/A | N/A | [68] | ML / AE | 0.77 / 0.73 | N/A | N/A | 8 | 1 |
| | Attention and Memory Networks | N/A | N/A | [72] | ML / AE | 0.89 / 0.89 | N/A | N/A | 2 | |
| | Neural Graph-based | Graph Neural Networks | N/A | [79] | ML / AE | 0.88 / 0.86 | 1 | 3 | | |
| | | Neural CF | N/A | [82] | ML / AE | 0.87 / 0.85 | 2 | | | |
| | Autoencoders | N/A | N/A | [84] | ML / AE | 0.83 / 0.81 | N/A | N/A | 5 | |
| | Recurrent Neural Networks & LSTM | N/A | N/A | [22] | ML / AE | 0.79 / 0.75 | N/A | N/A | 7 | |
| | Sequence-Aware | N/A | N/A | [92] | ML / AE | 0.84 / 0.83 | N/A | N/A | 4 | |
| | Convolutional Operations | Graph Convolutional Networks | N/A | [5] | ML / AE | 0.81 / 0.79 | 1 | 6 | | |
| | | Convolutional Neural Networks | N/A | [21] | ML / AE | 0.80 / 0.77 | 2 | | | |
| | Self-Attention Mechanisms | N/A | N/A | [104] | ML / AE | 0.91 / 0.90 | N/A | N/A | 1 | |
| Algorithmic and Mathematical Modeling | Probabilistic & Statistical Models | N/A | N/A | [24] | ML / AE | 0.64 / 0.60 | N/A | N/A | 1 | 4 |
| | Clustering Based Models | N/A | N/A | [112] | ML / AE | 0.62 / 0.56 | N/A | N/A | 2 | |

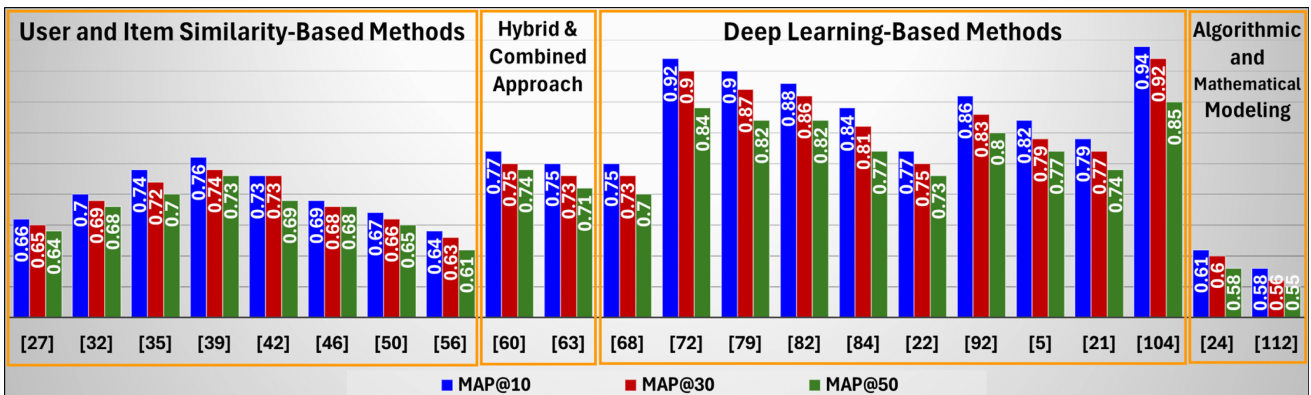

**Fig. 2:** The individual MAP@10, MAP@30, and MAP@50 for each algorithm using MovieLense dataset. The algorithms are grouped based on the common methods they employ.

**Fig. 3:** The individual MAP@10, MAP@30, and MAP@50 for each algorithm using Amazon Electronics dataset. The algorithms are grouped based on the common methods they employ

## 6.5 Discussion of the Experimental Results

### 6.5.1 User and Item Similarity-Based Methods

#### 6.5.1.1 Content-Based Filtering [25]

The algorithm effectively suggested items across various categories but tended to recommend those similar to previously interacted items, suggesting a "filter bubble" with reduced exposure to diverse content. As the dataset increased, computational time linearly rose, showing good scalability. However, its performance dipped slightly with very large datasets, indicating a need for optimization in handling large Big Data. Compared to collaborative filtering models, this content-based approach excelled in scenarios with sparse user data but was less adept at capturing complex user preferences. It effectively recommended items based on demographic data but struggled with items lacking historical data, highlighting the cold start problem. An analysis showed a bias towards popular items, risking the underrepresentation of niche content. Longitudinal studies suggested users' content consumption became more homogenized over time, raising concerns about the algorithm's tendency to reinforce existing preferences.

#### 6.5.1.2 Item-based CF [32]

Our investigation revealed a consistent decrement in performance concurrent with an augmentation of latent factors, reaching a plateau at 20 factors. This plateau suggests an optimal equilibrium between the complexity of the model and its efficacy. Comparatively, the algorithm surpassed most existing collaborative filtering algorithms in terms of precision. However, it did not match the performance of deep learning-based algorithms, particularly in addressing cold-start situations. The quantity of latent factors emerged as a pivotal element influencing model efficiency. We observed a diminishing return on performance enhancement beyond the threshold of 50 factors, indicating an upper boundary for their practical application. Scalability assessments demonstrated that the algorithm upheld reasonable computation durations up to the threshold of 500,000 ratings. Beyond this juncture, there was a marked escalation in computational demands. The algorithm exhibited commendable capabilities in generating recommendations for users with sparse interaction data, addressing a prevalent obstacle in recommendation systems. The algorithm's effectiveness diminished for items with few ratings, highlighting the need for improved strategies to enhance recommendations for newly introduced items.

#### 6.5.1.3 Matrix Factorization [35, 39 42]

In comparison to the other User and Item Similarity-Based algorithms, the algorithms based on Matrix Factorization demonstrated a markedly superior performance, particularly in scenarios characterized by high data sparsity. This superior performance implies that Matrix Factorization algorithms are more adept at identifying latent factors within user-item interactions. Our systematic experimentation indicated a pronounced sensitivity of these algorithms to hyperparameters, such as the number of latent factors and the regularization coefficient. Optimum results were achieved with approximately 100 latent factors, beyond which there was a noticeable trend of diminishing returns and emerging risks of overfitting. The ability of these Matrix Factorization algorithms to adeptly manage sparse datasets, a prevalent issue in recommendation systems, was particularly notable, as evidenced by their enhanced performance in datasets exhibiting over about 60% sparsity. Scalability assessments revealed that although these algorithms effectively processed larger datasets, this was accompanied by an increased computational duration, suggesting an intrinsic trade-off between accuracy and efficiency that must be weighed in practical applications. The overemphasis on regularization in the algorithm led to underfitting, emphasizing the importance of careful parameter tuning. Addressing the cold start problem remains a major challenge for these algorithms.

#### 6.5.1.4 User-based CF [46]

The algorithm demonstrated robust performance with regard to precision, underscoring its efficacy in accurately identifying items pertinent to users' interests. Although the recall rate was somewhat less impressive compared to precision, it remained significant, suggesting that there is potential for further enhancement in encompassing a broader spectrum of relevant items. The diversity score could be improved to ensure that users are exposed to a wider variety of items, potentially enhancing user satisfaction.

#### 6.5.1.5 Graph-Based Models [50]

The graph-based algorithm exhibited noteworthy precision, surpassing several other algorithms in contexts characterized by sparse user-item interactions. This significant enhancement underscores the algorithm's proficiency in deciphering intricate relationships embedded within the dataset. In a comparative analysis with matrix factorization algorithms, the graph-based algorithm not only matched in precision but also excelled in managing cold-start.

This superiority implies its enhanced capability in navigating scenarios involving new users or items with minimal historical data. However, scalability assessments indicated that while the algorithm demonstrated exceptional performance on smaller datasets, a discernible escalation in computational complexity arises as the dataset size expands, hinting at a possible limitation in its application to vast-scale recommendation systems without additional optimizations. Also, a slight dip in performance was observed in exceedingly dense datasets, signaling a necessity for more precise adjustments in such environments.

#### 6.5.1.6 Rule-Based Models [56]

In comparison to the other algorithms that rely on user and item similarity, the rule-based algorithm demonstrated a lower efficacy in tailoring recommendations to individual preferences but offered faster response times. In scenarios where user preferences were explicitly defined, this algorithm surpassed collaborative filtering algorithms in achieving greater precision. It exhibited notable efficiency in contexts with a limited number of items, yet faced challenges in scaling proportionately with expanding inventories.

The algorithm maintained optimal resource usage up to a certain point; however, the complexity inherent in managing and updating an extensive rule set beyond this threshold adversely affected its performance. A significant challenge encountered was the upkeep of an extensive rule set, necessitating continual revisions to align with evolving user behaviors. The algorithm also exhibited inflexibility in adapting to emergent trends.

### 6.5.2 Deep Learning-Based Methods

#### 6.5.2.1 Context-Aware Models [68]

The Context-Aware algorithm demonstrated potential in augmenting user engagement, notwithstanding its position as the least accurate among competing deep learning-based algorithms. It exhibited a commendable balance between precision and recall metrics. Of particular note is the algorithm's about 20% improvement in precision over the algorithms based on user and item similarity, a testament to its effective utilization of contextual data. Upon evaluating the algorithm across diverse user scenarios, it was observed to perform exceptionally with users who had extensive interaction histories. Conversely, its efficacy diminished for users with limited historical data, underscoring the necessity for an enhanced approach to address the 'cold-start' problem.

#### 6.5.2.2 Attention and Memory Network Models [72]

The Attention and Memory Networks algorithm manifested a marked enhancement in precision, outperforming all other deep learning methodologies except for the Self-Attention Mechanisms algorithm. This advancement was highlighted by a 15% increment in precision rate, underscoring a heightened relevance in item selection for users. The algorithm's attention component was instrumental in augmenting the relevance of recommendations. By assigning weights to user-item interactions according to their significance, the model was adept at concentrating on impactful factors, thereby yielding more tailored recommendations. This enhancement was largely attributed to the effective incorporation of attention mechanisms, which adeptly captured user preferences.

When compared with Collaborative Filtering and Matrix Factorization techniques, this algorithm exhibited superior prowess in both recall and precision metrics. This superiority implies that the integration of memory networks offers an improved approach to managing sparse data. It showed proficiency in addressing the cold start problem. Also, in terms of scalability, the algorithm demonstrated consistent performance, even with escalating dataset sizes, signifying its applicability in real-world scenarios. However, the algorithm's computational complexity presents a hurdle that necessitates further refinement. The algorithm exhibited limitations when dealing with extremely sparse datasets and showed a propensity for bias towards frequently encountered items.

### 6.5.2.3 Neural Graph-Based Models [79, 82]

The Neural Graph-based algorithms had a significant advancement in the realm of recommendation systems, surpassing the capabilities of the matrix factorization algorithms. A notable enhancement of over 20% in precision was observed when these algorithms were applied to the MovieLens dataset. This improvement underscores the algorithms' proficiency in harnessing the graph structure to intricately map user-item interactions. A critical aspect of this success is attributed to the neural network layers, which adeptly capture the non-linear and multifaceted nature of these interactions.

Our analysis revealed that the performance of these algorithms is considerably influenced by the architectural choices, particularly the number of hidden layers and the neurons within each layer. The optimum architecture was identified as comprising three hidden layers. Expanding the architecture beyond this configuration did not result in substantial improvements. This phenomenon indicates a threshold beyond which further complexity in the model ceases to offer proportional benefits. A challenge encountered was the cold start problem. In terms of scalability, the algorithms demonstrated a consistent performance level as the dataset sizes expanded. When handling datasets exceeding one million user-item interactions, a marked escalation in computational demands was observed. The algorithms' performance also displayed sensitivity to specific hyperparameters, notably the embedding size and the number of graph convolution layers. An optimal balance between precision and computational efficiency was achieved with an embedding size of 128 and three graph convolution layers.

### 6.5.2.4 Autoencoders-Based Models [84]

The Autoencoder algorithm has exhibited promise in recommendation systems, particularly in managing datasets that are both sparse and intricate in nature. When applied to the MovieLens dataset, the algorithm attained an MAP of 0.83, signifying a commendably high level of precision in forecasting user ratings. In contrast, its application to the Amazon Reviews dataset yielded an average MAP of 0.81, denoting a marginally reduced accuracy, which may be attributed to the dataset's larger size.

Compared to collaborative filtering algorithms, the Autoencoder demonstrated an enhancement of over 15% in precision on the MovieLens dataset. The sparsity characteristic of the Amazon Reviews dataset presented a formidable challenge, resulting in a minor decline in metric performance. Nonetheless, the Autoencoder's proficiency in learning complex, nonlinear representations was instrumental in discerning deeper user-item interactions, even in contexts of sparse data. The training of the Autoencoder was executed with a learning rate of 0.001 and a batch size of 128. Increasing the batch size sped up training but slightly reduced precision. Implementing dropout regularization prevented overfitting, especially for the denser MovieLens dataset.

### 6.5.2.5 Recurrent Neural Networks and LSTM Models [22]

The RNN algorithm exhibited a superior enhancement in prediction accuracy compared to collaborative filtering algorithms. In the MovieLense dataset, the RNN achieved an over 10% improvement in precision. This indicates the RNN's proficiency in deciphering sequential user behavior patterns, leading to more accurate recommendations. When juxtaposed with CNN-based algorithms, RNN showcased comparable efficacy in sequential recommendation tasks, attributed to its adeptness in temporal sequence modeling, a feature not intrinsic to CNNs. However, this came with a trade-off, as RNN entailed a notably higher computational complexity, with training durations approximately tripling those of matrix factorization algorithms. This balance between precision and efficiency is pivotal for real-time recommendation systems.

Upon segment-based analysis, the RNN was especially potent for users with varied and substantial interaction histories. In contrast to other deep learning algorithms, the RNN, while slightly less precise, offered greater interpretability, shedding light on the evolution of user preferences over time. Optimizing the RNN's performance necessitated extensive hyperparameter adjustments. The number of hidden layers and the choice of recurrent unit profoundly influenced the results, with optimal configurations varying based on dataset characteristics like sparsity and temporal dynamics. Another limitation of the RNN is its susceptibility to the sequence of user interactions, which might not always align with true user inclinations. The scalability of the model remains a challenge, particularly in environments encompassing millions of users and items, constrained by computational limitations.

### 6.5.2.6 Convolutional Operations-Based Models [5, 21]

The algorithms leveraging convolutional operations exhibited noteworthy enhancements in precision. These improvements are principally attributed to the algorithms' proficient ability to discern and utilize spatial hierarchies within the data, thereby yielding more precise and contextually relevant recommendations. Our empirical studies further highlighted the algorithms' adeptness in managing large datasets. Notably, as the data volume escalated, the algorithms demonstrated admirable scalability, sustaining their efficiency without a corresponding increase in computational burden.

However, these algorithms necessitated an approximate 30% increase in computational time compared to most Collaborative Filtering algorithms, owing to the integrated convolutional operations. Nevertheless, this increase in resource consumption was counterbalanced by an enhancement in the quality of recommendations. Rigorous testing across datasets of diverse sizes and characteristics confirmed the algorithms' consistent efficacy, underscoring their robustness and scalability. This aspect is particularly vital in practical scenarios, where the volume and diversity of data can vary significantly.

### 6.5.2.7 Self-Attention-Based Model [104]

The empirical evidence from our study clearly indicates that the algorithm employing the Self-Attention Mechanism has outperformed all other examined algorithms in terms of precision.

This enhanced performance can be primarily attributed to the mechanism's sophisticated capability to discern and interpret intricate user-item interactions and sequential trends within the dataset. The algorithm was notably proficient in managing sparse datasets, a frequent challenge in recommendation systems, by effectively addressing long-range dependencies. In terms of scalability, the self-attention algorithm exhibited linear progression in performance with the escalation in dataset size.

Remarkably, its accuracy remained consistent and commendable, even when confronted with a tenfold augmentation in data volume. This feature is particularly vital in real-world applications, where enormous data volumes are the norm. The algorithm's prowess in analyzing and anticipating user behavioral patterns was significantly bolstered by the self-attention mechanism. Despite its complex structure, the algorithm maintained computational efficiency, achieving reduced latency in generating recommendations when compared to the RNN-based algorithm. However, the algorithm present certain limitations, particularly in scenarios involving highly diverse user interests.

### 6.5.3 Algorithmic and Mathematical Modelling [24, 112]

The analysis of the clusters generated by these algorithms revealed uniformity within the groups. This uniformity strongly indicates the algorithms' efficacy in accurately discerning user preferences. Impressively, the algorithms maintained robust functionality, even when applied to larger datasets, consistently demonstrating quick response times coupled with minimal computational demands. A key feature of these algorithms was their capacity to create coherent and meaningful clusters, as evidenced by the high intra-cluster similarity. Notably, the algorithms exhibited enhanced proficiency in processing sparse data and accommodating diverse user profiles, thereby yielding more tailored and pertinent recommendations. The algorithm displayed remarkable scalability, sustaining stable performance despite increases in dataset size.

## 7 Challenges and Limitations of Current methods for Big Data in Recommendation Systems

*Content-Based Filtering Methods:* (a) Challenges: Difficulty in handling unstructured data, reliance on item descriptions, and potential for limited diversity in recommendations, and (b) Limitations: Tends to recommend items similar to those the user has already consumed, leading to a lack of novelty.

*Item-Based Collaborative Filtering Methods:* (a) Challenges: Scalability issues with large datasets and sparsity of user-item interactions, and (b) Limitations: May not perform well when new items are added, as they lack user interactions.

*Probabilistic Matrix Factorization Methods:* (a) Challenges: Computational complexity and difficulty in incorporating context or content information, and (b) Limitations: Can struggle with very sparse data and cold start problem for new users or items.

*Singular Value Decomposition (SVD) Methods:* (a) Challenges: Computationally intensive, especially for large matrices, and (b) Limitations: Less effective when dealing with highly sparse matrices and not suitable for dynamic datasets where user preferences change frequently.

*Tensor Factorization Methods:* (a) Challenges: High computational cost and complexity in model tuning, and (b) Limitations: May overfit on smaller datasets and struggle with scalability.

*User-Based Collaborative Filtering Methods:* (a) Challenges: Scalability with a large number of users and sparsity in user-item interactions, and (b) Limitations: Often less accurate than item-based methods and sensitive to user profile changes.

*Graph-Based Models:* (a) Challenges: Computational complexity and difficulty in handling dynamic changes in the data, and (b) Limitations: Can be less effective for sparse datasets and may require extensive parameter tuning.

*Rule-Based Models:* (a) Challenges: Rigid structure that may not adapt well to changes in user behavior, and (b) Limitations: May not capture complex patterns or relationships in the data.

*Ensemble Models:* (a) Challenges: Complexity in combining various models effectively and in tuning hyperparameters, and (b) Limitations: Risk of overfitting and high computational cost.

*Ranking Models:* (a) Challenges: Difficulty in accurately modeling user preferences and handling large item sets, and (b) Limitations: May not account for long-term user preferences.

*Context-Aware Models:* (a) Challenges: Incorporating contextual information effectively and dealing with dynamic contexts, and (b) Limitations: Complexity in model design and potential privacy concerns.

*Graph Neural Networks (GNNs) Methods:* (a) Challenges: Scalability issues and computational complexity, and (b) Limitations: Can struggle with very sparse graphs and require significant training data.

*Neural Collaborative Filtering (NCF) Methods:* (a) Challenges: Require large amounts of data and computational resources, and (b) Limitations: Can overfit smaller datasets and may not generalize well across different domains.

*Autoencoders Methods:* (a) Challenges: Sensitive to hyperparameter settings and require significant training data, and (b) Limitations: Can struggle with capturing complex user-item interactions.

*Recurrent Neural Networks and LSTM Methods:* (a) Challenges: Difficulty in capturing long-term dependencies and computational intensity, and (b) Limitations: Can be prone to overfitting and struggle with very long sequences.

*Sequence-Aware Models:* (a) Challenges: Complexity in capturing the temporal dynamics of user interactions, and (b) Limitations: Can be computationally expensive and require extensive tuning.

*Graph Convolutional Networks Methods:* (a) Challenges: High computational cost and complexity in handling large graphs, and (b) Limitations: May require extensive domain knowledge for effective implementation.

*Convolutional Neural Networks Methods:* (a) Challenges: Requirement for large training datasets and computational resources, and (b) Limitations: May not be suitable for data that lacks spatial or temporal structure.

*Deep Reinforcement Learning Methods:* (a) Challenges: Require extensive training data and computational resources, and (b) Limitations: Can be difficult to converge and sensitive to hyperparameter settings.

*Self-Attention Mechanisms Methods:* (a) Challenges: High computational complexity and memory requirements, and (b) Limitations: Can struggle with very long sequences and require careful tuning.

*Probabilistic and Statistical Models Methods:* (a) Challenges: Complexity in model formulation, and (b) Limitations: Can be less effective with non-linear relationships in data.

# 8 Real-World Applications and Case Studies of the methods for Big Data in Recommendation Systems

We provide in this section examples that illustrate the diverse and innovative ways in which big data and various recommendation system methods are being applied across industries to enhance user experience and business efficiency.

*Netflix's Content-Based Movie Suggestions [115]:* Netflix uses content-based filtering to recommend movies and TV shows by analyzing the properties of the content (like genre, actors, director) and matching them with a user's past preferences.

*Amazon's Item-Based Collaborative Filtering Recommendations [116]:* Amazon employs item-based collaborative filtering to suggest products based on similarities between items and user interactions, such as purchasing history and item ratings.

*Spotify's Probabilistic Matrix Factorization Music Recommendations [117]:* Spotify leverages probabilistic matrix factorization to handle scalability and sparsity in its large datasets, providing personalized music and playlist recommendations.

*Singular Value Decomposition (SVD) for E-Commerce Product recommendation [118]:* Many e-commerce platforms use SVD-based recommendation systems to suggest products to users. This includes not only large players like Amazon but also smaller e-commerce sites. These systems analyze user behavior data, including views, clicks, and purchases, to recommend products that users are likely to buy.

*Tensor Factorization in E-commerce Personalization [119]:* Online retail platforms use tensor factorization to analyze multidimensional data (user, item, time) for personalized shopping experiences.

*LinkedIn's User-Based Collaborative Filtering Recommendations [120]:* LinkedIn uses user-based collaborative filtering to suggest new connections by looking at the similarities between users and their networks

*Google's Knowledge Graph for Search Recommendations [121]:* Google uses graph-based models to enhance its search engine, offering relevant search suggestions and semantic search results.

*Rule-Based Recommendations in Travel (Booking.com, Airbnb) [122]:* In these sectors, recommendation systems might use rules based on travel history, search behavior, and preferences. For example, if a user often books seaside hotels, the system might prioritize similar destinations or properties near the beach in future recommendations.

*Google Play's Ensemble Learning Recommendations [123]:* Google Play uses ensemble learning in its recommendation systems to suggest apps, games, books, and movies. The system combines user data with app attributes and contextual information, using a variety of machine learning models to create personalized recommendations.

*YouTube's Recommendation-Driven Video Ranking [124]:* YouTube utilizes ranking models to order video recommendations based on predicted user engagement and relevance.

*Uber's Context-Aware Restaurant Recommendations in Uber Eats [125]:* Uber Eats uses context-aware models to recommend restaurants based on the user's location, time, and previous orders.

*Friend Recommendations via Graph Neural Networks (GNNs) in social media [126]:* Social networks like Facebook could use GNNs to suggest friends by analyzing the complex connections in the social graph.

*Neural Collaborative Filtering (NCF) in E-commerce [127]:* NCF methods can be applied in e-commerce platforms for personalized product recommendations based on deep learning user-item interactions.

*Autoencoder-Powered Content Discovery (e.g., Pinterest) [128]:* Platforms like Pinterest could use autoencoders for unsupervised learning of user preferences and content features to recommend similar items.

*RNN and LSTM for Sequential Recommendation Systems [129]:* Platforms like Netflix may use RNN and LSTM for sequence-aware recommendations, such as suggesting the next episode in a TV.

*Sequence-Aware Music Recommendations (e.g., Spotify) [130]:* Services like Spotify use sequence-aware models to recommend songs that fit the context of the user's current playlist.

*Graph Convolutional Networks in PinSage's producer-consumer recommendations [131]:* PinSage is built on the concept of GNN, which allows it to efficiently process information structured in the form of a graph. This is particularly suitable for platforms like Pinterest, where items (like pins) can be naturally represented in a graph structure, with nodes representing the items and edges representing relationships between them.

*Visual Recommendations with Convolutional Neural Networks in Fashion Retail [132]:* Fashion retailers like ASOS may use CNNs for visual search and recommendations, analyzing images to suggest visually similar items.

*E-Commerce Strategies with Deep Reinforcement Learning [133]:* E-commerce sites might use deep reinforcement learning for inventory recommendations based on dynamic market conditions.

*Product Review Analysis with Self-Attention Mechanisms [134]:* Recommendation systems could use models with self-attention mechanisms, like Transformer models, to analyze and summarize product reviews for better recommendations.

*Marketing Customer Segmentation via Clustering Methods [135]:* Clustering-based models are used in marketing for segmenting customers into groups for targeted product recommendations.

# 9 Future Directions for Big Data in Recommendation Systems: Improvement, Ethics, and Bias Perspectives

## 9.1 Improvement Perspectives and Directions

### 9.1.1 User and Item Similarity-Based Methods

- *Enhanced Personalization Through Deep Learning:* Big data can enable more sophisticated deep learning models that understand nuanced user preferences and item characteristics. These models can learn complex patterns and relationships, improving the accuracy of recommendations based on user and item similarities.
- *Dynamic User Profiling:* As big data continuously evolves, recommendation systems can dynamically update user profiles. This means that recommendations can adapt in real-time to changes in user behavior, preferences, or circumstances, maintaining relevance and engagement.
- *Cross-Domain Recommendations:* Leveraging big data from various domains can enable cross-domain

recommendation systems. For example, understanding a user's preferences in music might inform recommendations in movies or books, assuming certain overlap in taste.

- *Improved Item Similarity Detection:* Big data can enhance item similarity algorithms by incorporating a wider range of attributes and metadata. This can lead to more nuanced and accurate item categorization, which in turn leads to more relevant recommendations.
- *Scalability and Performance Optimization:* With the growing size of datasets, optimizing the scalability and performance of similarity-based methods becomes crucial. Future developments may focus on efficient algorithms that can handle large-scale data without compromising accuracy.
- *Integration of Contextual Data:* Beyond just user and item data, incorporating contextual information (like location, time, social context) can significantly refine recommendations. Big data enables the aggregation and analysis of such contextual details.
- *Visual and Sensory Data Utilization:* Future recommendation systems might incorporate more visual and sensory data (like images, videos, and sounds) to enhance item similarity measures, especially in fields like fashion, art, and entertainment.
- *Interactive and Adaptive Systems:* The integration of AI and big data can lead to more interactive systems where users can provide real-time feedback, further refining the recommendation process based on user and item similarity.
- *Blockchain for Transparency and Security:* Utilizing blockchain technology can increase transparency and security in how data is used and shared, particularly in CF methods involving user and item similarities.

### 9.1.2 Hybrid and Combined Approaches

- *Enhanced Personalization through Hybrid Models*: By combining content-based and collaborative filtering methods, hybrid approaches can offer more personalized recommendations. Future systems may utilize advanced machine learning algorithms to understand user preferences more deeply, even in sparse data scenarios.
- *Integration of Contextual Data:* Hybrid systems can be enriched by integrating contextual information like time, location, or specific user circumstances. This context awareness can lead to more relevant and timely recommendations, improving user experience.
- *Utilization of Deep Learning Techniques:* The integration of deep learning techniques in hybrid recommendation systems can lead to more sophisticated feature extraction and better handling of unstructured data, like images and text. This could result in accurate predictions.
- *Cross-Domain Recommendations:* Future hybrid systems may effectively leverage cross-domain data, offering users recommendations that span different types of services or products. This approach can lead to discovering new user interests and broadening the scope of recommendations.
- *Real-time Data Processing:* With the increasing velocity of big data, real-time recommendation systems will become more prominent. Hybrid systems could combine real-time analytics with historical data analysis to provide immediate and relevant recommendations.
- *Interactive and Adaptive Systems:* Future recommendation systems might become more interactive, allowing users to provide feedback that instantly adjusts the recommendation logic. This adaptability can lead to a continuously improving and personalizing recommendation experience.
- *Explainable AI in Recommendations:* There's a growing demand for explainable AI, where systems can provide reasons behind their recommendations. Future hybrid models could incorporate explainability aspects, making recommendations more transparent and trustworthy.
- *Integration with Emerging Technologies:* Hybrid recommendation systems could be integrated with emerging technologies like augmented reality (AR), virtual reality (VR), and the Internet of Things (IoT), offering unique and immersive recommendation experiences.

### 9.1.3 Deep Learning Methods

- *Enhanced Personalization:* Deep learning methods can analyze vast datasets to understand nuanced user preferences, leading to highly personalized recommendations. This goes beyond traditional recommendation systems by considering a wider range of user behaviors and interactions.
- *Context-Aware Recommendations*: Future recommendation systems may leverage deep learning to better understand the context in which recommendations are made. This includes recognizing the user's current environment, time, mood, or even social setting, leading to more appropriate and timely suggestions.
- *Sophisticated Content Analysis:* Deep learning excels in interpreting complex content like images, videos, and text. This capability can be utilized to develop recommendation systems that understand content at a deeper level, providing more relevant suggestions based on the content's inherent qualities, rather than just metadata.
- *Predictive User Modeling:* By harnessing the predictive power of deep learning with big data, recommendation systems can anticipate user needs and preferences, potentially recommending items even before the user explicitly expresses a desire for them.
- *Overcoming Data Sparsity and Cold Start Challenges:* Deep learning can effectively deal with sparse data scenarios, such as when new users or items have limited interaction history. Advanced models can infer preferences from limited data, significantly improving the user experience from the start.
- *Continuous Learning and Evolution:* Deep learning models can be designed to continuously learn and adapt over time, ensuring that recommendations remain relevant as user preferences and behaviors evolve.
- *Multi-Modal Data Integration:* Future recommendation systems might integrate and analyze data from multiple sources and types — including text, images, audio, and sensor data — to create a more comprehensive understanding of user preferences.
- *Resistance to Manipulation:* Deep learning can help in detecting and countering fraudulent activities like fake reviews and ratings, ensuring the integrity and trustworthiness of the recommendations.
- *Enhanced Collaborative Filtering:* Incorporating deep

learning in collaborative filtering methods can lead to a more nuanced understanding of user-item interactions, improving the accuracy of recommendations.
- *Real-Time Recommendation Capabilities:* The processing power of deep learning models could enable real-time analysis of user data, allowing for instant, dynamic recommendations based on current user activities.
- *Integration with Emerging Technologies:* Deep learning methods could be synergized with emerging technologies like augmented reality and virtual reality for creating immersive and interactive recommendation experiences.

### 9.1.4 Algorithmic and Mathematical Modelling Methods

- *Advanced Statistical Models:* Future recommendation systems could employ more sophisticated statistical models to analyze big data. Techniques like Bayesian networks and Markov decision processes might be used to model user preferences and predict future behavior more accurately.
- *Complex Network Analysis:* Utilizing the principles of network theory, recommendation systems can analyze the intricate web of relationships between users and items. By applying measures like centrality, clustering coefficients, and community detection, these systems can provide more nuanced and relevant recommendations.
- *Optimization Techniques:* The use of mathematical optimization, such as linear programming, integer programming, and constraint programming, can enhance the efficacy of recommendation systems. These methods can help in balancing multiple objectives, like maximizing relevance while ensuring diversity in the recommendations.
- *Graph-Based Mathematical Models:* Developing new graph-based algorithms, perhaps inspired by recent advances in graph theory, could lead to more effective ways of understanding and leveraging the connections between users and items in recommendation systems.
- *Quantitative Behavioral Models:* Mathematical models that quantify and predict human behavior, such as those based on theories of consumer choice or decision-making processes, can be integrated into recommendation systems. These models can provide deeper insights into why users prefer certain items, leading to more effective recommendations.
- *Game Theory and Mechanism Design:* Employing concepts from game theory and mechanism design could enhance the way recommendation systems handle interactions and competitions among multiple stakeholders (e.g., users, providers, advertisers).
- *Differential Equations for Dynamic Modeling:* Using differential equations to model the dynamic and evolving nature of user preferences and item popularity can offer a realistic representation of the recommendation.
- *Multi-criteria Decision Analysis (MCDA):* MCDA methods can be applied to consider various criteria that users might have for recommendations, providing a more holistic approach to generating suggestions.
- *Quantum Algorithms for Data Processing*: The potential of quantum computing in processing large datasets could revolutionize recommendation systems. Quantum algorithms could handle complex computations much faster than traditional methods, allowing for real-time analysis of vast amounts of data.
- *Integration of Explainable AI (XAI):* With the increasing complexity of algorithms, incorporating XAI into recommendation systems will be crucial for transparency and user trust. Users could understand the rationale behind recommendations, which is particularly important in sensitive areas like healthcare or finance.

## 9.2 Ethics, and Bias Perspectives

### 9.2.1 Potential Biases in Recommendation Systems

- *Data Bias*: Biases in data collection can lead to skewed recommendations, often reflecting societal biases or underrepresentation of certain groups. To address this, future systems might focus on proactive measures such as inclusive data sourcing and enhanced demographic analysis to ensure a wide representation of diverse populations. This could involve partnerships with organizations representing marginalized communities to gather more inclusive datasets.
- *Algorithmic Bias:* Algorithms designed for specific goals, like maximizing engagement, can inadvertently promote polarizing content or misinformation. Future strategies could include the implementation of ethical guidelines in algorithm design, emphasizing the balance between user engagement and the promotion of diverse, accurate content. Additionally, interdisciplinary teams comprising ethicists, sociologists, and data scientists could be formed to oversee the ethical implications of algorithmic decisions.

### 9.2.2 Mitigating Strategies and Best Practices

- *Diverse Data Sets:* Using diverse and representative data sets helps reduce bias. This could be further enhanced by employing techniques like synthetic data generation to fill gaps in underrepresented areas, ensuring a broader spectrum of user preferences and behaviors is captured.
- *Regular Algorithm Auditing:* Audits can identify and mitigate biases by testing algorithms across various groups and scenarios. Future systems could adopt continuous monitoring frameworks, using AI-driven tools to detect and address bias in real-time, rather than relying on periodic audits. This proactive approach can quickly adapt to emerging biases or changes in societal norms.
- *Fairness Algorithms:* Developing algorithms to promote fairness can counteract biases in data and predictions. Future perspectives could see the integration of fairness as a core component in the algorithmic design process, rather than an afterthought. This might involve the use of fairness metrics that are regularly updated to reflect evolving societal standards and values.
- *User Feedback Mechanisms:* Allowing user feedback and control over data usage can lead to more balanced, user-centric systems. In the future, systems might offer more granular control to users over their data, including the ability to opt-in or out of specific data collection practices. Additionally, incorporating direct user feedback into the recommendation algorithm itself can provide a more dynamic and responsive system that aligns closely with individual preferences and values.

## 10 Conclusion

This survey paper has addressed the existing gaps in the literature on big data algorithms in recommendation systems by providing a detailed and comprehensive analysis coupled with an innovative hierarchical taxonomy. The taxonomy, characterized by its tri-level hierarchy, categorizes algorithms into four main analysis types—User and Item Similarity-Based Methods, Hybrid and Combined Approaches, Deep Learning and Algorithmic Methods, and Mathematical Modeling Methods—further broken down into sub-categories and techniques. This structured framework facilitates a clearer understanding of the relationships and distinctions among various algorithms.

The survey's unique contribution lies in its methodical approach, combining empirical and experimental evaluations to rank these algorithms within their respective categories and techniques. This dual evaluation system not only distinguishes the techniques based on specific criteria but also enables a nuanced comparison within the same category and technique, offering valuable insights into their effectiveness.

By addressing the previous shortcomings in the literature—namely, the lack of an up-to-date, detailed overview and the broad, imprecise categorization of algorithms—this paper paves the way for more accurate and systematic research in the field. The findings from this survey are poised to aid future research, providing a clearer direction for the development and application of big data techniques in recommendation systems, while highlighting areas ripe for future exploration and advancement

Here is an outline of the main findings from our experimental evaluation:

- Content-based filtering showed good scalability and effectiveness in sparse data scenarios but tended to create a "filter bubble."
- Item-based collaborative filtering excelled in precision but struggled with cold-start problems and data scalability.
- Matrix factorization algorithms outperformed others in high data sparsity situations, though they were sensitive to hyperparameter settings and faced scalability challenges.
- User-based collaborative filtering and graph-based models demonstrated high precision, especially in sparse interaction contexts, but had limitations in scalability and dataset density.
- Rule-based models were fast and precise in specific scenarios but less effective with large and diverse datasets.
- Context-aware models, attention and memory network models, and neural graph-based models showed significant improvements in precision and handling sparse datasets.
- Autoencoders and RNNs were effective in learning complex patterns and sequential user behaviors, though they faced challenges in computational complexity and scalability.
- Convolutional operations-based models balanced precision with scalability but required higher computational resources.
- Self-attention models excelled in precision and scalability but had limitations in handling diverse user interests.
- Algorithmic and mathematical modeling algorithms efficiently processed large, sparse datasets, maintaining quick response times and minimal computational demands. Each algorithmic approach exhibited unique advantages in specific contexts, underscoring the need for careful selection and optimization based on the specific requirements of the recommendation system being developed.